\documentclass[useAMS,usenatbib]{mn2e}
\usepackage{mathrsfs} 
\usepackage{graphicx} 
\usepackage{mediabb}
\usepackage{color} \usepackage{bm} 
\title
[CO and Radio Pillars of Creation in M16]
{CO-Line and Radio Continuum Study of Elephant Trunks: The Pillars of Creation in M16} 
\author[Yoshiaki Sofue]
{Yoshiaki \textsc{Sofue}\\
 Institute of Astronomy, The University of Tokyo, Mitaka, Tokyo 181-0015\\
 E-mail: sofue@ioa.s.u-tokyo.ac.jp} 

\begin{document} 
\date{ } 
\maketitle  
  
\def\vlsr{v_{\rm LSR}} \def\Msun{M_\odot} \def\Lsun{L_\odot} 
\def\deg{^\circ} \def\r{\bibitem[]{}} \def\/{\over} \def\kms{km s$^{-1}$} 
\def\Vsun{V_0} \def\Vrot{V_{\rm rot}} \def\Te{T_{\rm e}} \def\Tb{T_{\rm B}} 
\def\sin{{\rm sin}\ } \def\cos{{\rm cos}\ } \def\Hcc{ H cm$^{-3}$ } 
\def\co{$^{12}$CO$(J=1-0)$ }\def\coth{$^{13}$CO$(J=1-0)$ }
\def\coei{C$^{18}$O$(J=1-0)$ } 
\def\be{\begin{equation}} \def\ee{\end{equation}}
\def\Kkms{K \kms} \def\Ico{I_{\rm CO}} \def\Xco{X_{\rm CO}} 
\def\mH{m_{\rm H}} \def\x{\times}\def\Hcc{H cm$^{-3}$} 
\def\({\left(} \def\){\right)} \def\[{\left[} \def\]{\right]}  
\def\Kkms{ K \kms } \def\Hsqcm{ H cm$^{-2}$ } \def\hcc{{\rm H \ cm^{-3}}}
\def\mum{$\mu$m }    \def\Mvir{M_{\rm vir}}\def\htwocc{\htwo cm$^{-3}$ }
\def\htwosqcm{\htwo cm$^{-2}$ } \def\htwocol{\htwo cm$^{-2}$ } 
\def\htwo{H$_2$} \def\Htwo{H$_2$ } \def\NHtwo{N_{\rm H_2}} 
\def\nHtwo{n_{\rm H_2}} 
\def\Xco{X_{\rm CO}} \def\nHII{n_{\rm HII}} \def\nHtwo{n_{\rm H_2}} 
\def\Htwosqcm{H$_2$ cm$^{-2}$}\def\Tex{T_{\rm ex}}
\def\Ncoth{N_{\rm ^{13}CO}}\def\NHtwocoth{N_{\rm H_2}(^{13}{\rm CO})}
\def\nhtwo{n_{\rm H_2}}\def\Nhtwo{N_{\rm H_2}} \def\nel{n_{\rm e}}
\def\muG{$\mu$G} \def\Alf{Alfv{\'e}n }
\def\rev{ }

\begin{abstract} 
Molecular-line and radio continuum properties of the elephant trunks (ET, pillars of creation) in M16 are investigated by analyzing the \co, \coth and \coei-line survey data with the Nobeyama 45-m telescope and the Galactic plane radio survey at 20 and 90 cm with the Very Large Array. The head clump of Pillar West I is found to be the brightest radio source in M16, showing thermal spectrum and property of a compact HII region with the nearest O5 star in NGC 6611 being the heating source. The radio pillars have cometary structure concave to the molecular trunk head, and the surface brightness distribution obeys a simple illumination law by a remote excitation source. The molecular density in the pillar head is estimated to be several $10^4$ \htwocc and molecular mass $\sim 13-40 \Msun$. CO-line kinematics reveals random rotation of the clumps in the pillar tail at $\sim 1-2$ \kms, comparable to the velocity dispersion and estimated \Alf velocity. It is suggested that the random directions of velocity gradients would manifest a torsional magnetic oscillation of the clumps around the pillar axis. 
\end{abstract}  


\section{Introduction} 
Elephant trunks (ET) are cometary dark clouds with their head clumps being supposed to evolve into denser molecular globules to create stars. Typical examples are seen in the HII region M16 as the "pillars of creation" widely known by their Hubble Space Telescope (HST) images. ETs are considered to be produced by the Rayleigh-Taylor instability (RTI) at the interface of expanding HII region and molecular cloud
(Frieman 1954; Spitzer 1954; Osterbrock 1957; Macky and Lim 2010),
and/or radiation-driven implosion (DRI) by the pressure due to photo-dissociation of the cloud surface by UV photons from O stars
({Gritschneder et al.} {2010}; {Ercolano et al.} {2012}; {Haworth \& Harries} {2012}). 

A number of observations of ETs have been obtained from infrared to X rays, particularly extensively in the far infrared (FIR), in the decades
(
Hester et al. 1996; Pilbratt et al. 1998; {McCaughrean \& Andersen} {2002};
{Carlqvist, et al} {2002}; Urquhart et al. 2003;
Sugitani et al. 2007; {Linsky et al.} {2007} 
Chaughan et al. 2012; Getman et al. 2012; 
Hill et al. 2012; 
Schneider et al. 2016; 
M{\"a}kel{\"a} et al. {2017}; 
Pattle et al. 2018; Panwar et al. 2019 ). 
 See Oliveira (2008) for review on the ETs in M16 and star formation.

\begin{figure*} 
\begin{center}  
\includegraphics[width=14cm]{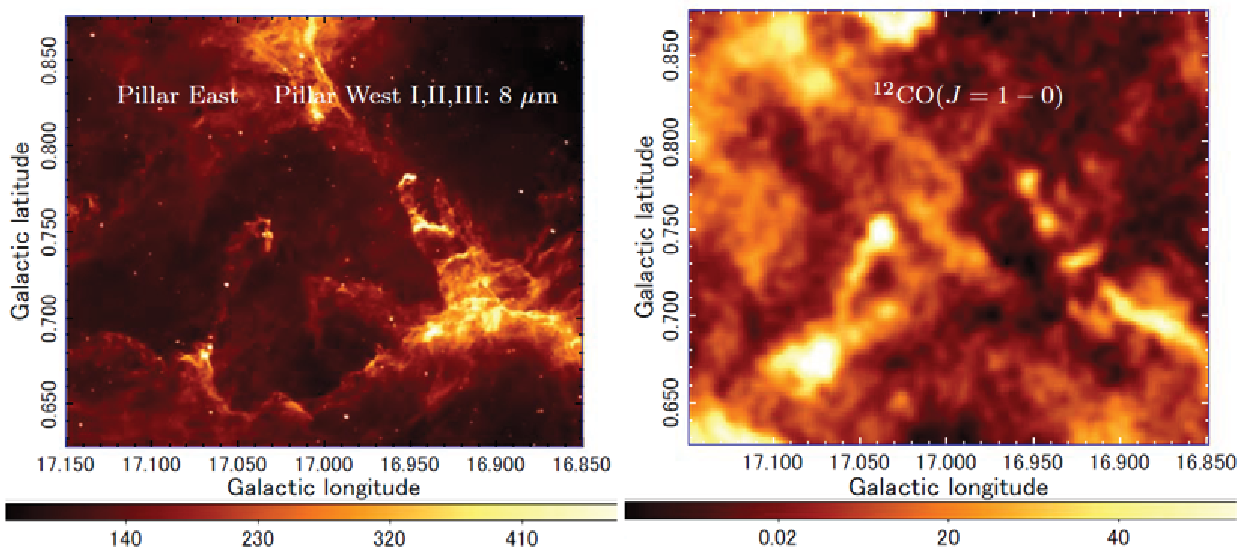}   
\end{center}
\caption{FIR $\lambda$ 8 \mum map  (Jy str$^{-1}$) (PAH (polycyclic aromatic hydrogen) emission from dust clouds' surface extracted from MAGPIS, and \co intensity  (\Kkms) map of the M16 region. Used nomenclature in this paper is indicated: Pillar East and Pillars West (pillars of creation) I, II, III).  } 
\label{bb_8_co}  
 	\end{figure*}

Molecular line observations in the CO lines, which directly measure the gaseous mass and kinematics,  have been also obtained in several ETs
({Sherwood \& Dachs} {1976}; 
 {Schneps et al.} {1980}; 
 {Gonzalez-Alfonso \& Cernicharo}{1994}; 
 {Massi et al.} {1997}; 
{Gahm et al.} {2006, 2013}; 
 {Haikala et al.} {2017} 
 {M{\"a}kel{\"a} et al.} 2017). 
 As to the ETs in M16, several CO-line observations
 have been obtained  ({Pound} {1998}; 
{White et al.} {1999}; 
Andersen et al. 2004;
 Schuller et al. 2006;
Pattle et al. 2018;  Xu et al. 2019).
In our recent paper on giant elephant trunks (GET) in the star forming spiral arms (Sofue 2019), we provisionally discussed the molecular property of ETs in M16 as a template. 
 Figure \ref{bb_8_co} reproduces the far-infrared (FIR) 8 \mum and \co line intensity maps from Sofue (2019) with the nomenclature used in this paper (Pillar East and Pillars West I, II and III), where the data have been extracted from the archival data bases described below.
  
In this paper, we perform a detailed analysis in CO and radio continuum of the prominent ETs at $l\sim 17\deg.04$ (Pillar East, or Eagle's pillar\footnote{https://apod.nasa.gov/apod/ap181202.html}), and at $l\sim 16\deg.95$ (Pillars West I, II and III, or, hereafter, Pillar I, II and III\footnote{https://apod.nasa.gov/apod/ap150107.html}). 
We adopt a distance of M16 to be $D=2.0$ kpc ({Hillenbrand et al.} {1993}; {Guarcello et al.} {2007}). 

We utilize the archival CO-line data base FUGIN (FOREST (FOur-beam REceiver System on the 45-m Telescope) Unbiased Galactic plane Imaging Survey at Nobeyama) (Umemoto et al. 2017; http://jvo.nao.ac.jp/portal/nobeyama/fugin.do), and  
the  FIR and radio continuum archival data of the Galactic Plane from the Multi-Array Galactic Plane Imaging Survey (MAGPIS) (Helfand et al. 2006; 
{\rev Churchwell et al. 2009}; https://third.ucllnl.org/gps/index.html).

\section{Data and maps}
        
\subsection{CO maps}
We first present the CO line data of the M16 region in the forms of channel maps, integrated intensity (moment 0), velocity field (moment 1), velocity dispersion (moment 2) maps, and longitude-velocity (LV) diagrams. The data were taken from the FUGIN survey (Umemoto et al. 2017). The observations were obtained in the molecular lines of $^{12}$CO, $^{13}$CO, C$^{18}$O ($J=1-0$) with the full beam width of half maximum of $14''$, $15''$ and $15''$. The angular resolutions on the resultant maps are $20''$, $21''$, and $21''$, respectively, which corresponds to 0.19, 0.20 and 0.20 pc. The velocity resolution was $\Delta v= 1.3$ \kms, and  the rms noise in the \co line data cube was $\delta T\sim 1.5$ K in average, which yielded noise in the integrated intensity of $\delta I \sim \delta T w \sqrt{w/\Delta v} \sim 3$ \Kkms, where $w\sim 4$ \kms is the line width. The scanning effect in the original data has been removed by applying the pressing method (Sofue and Reich 1979; Sofue 2019). 

\begin{figure*} 
\begin{center}  
\includegraphics[width=10cm]{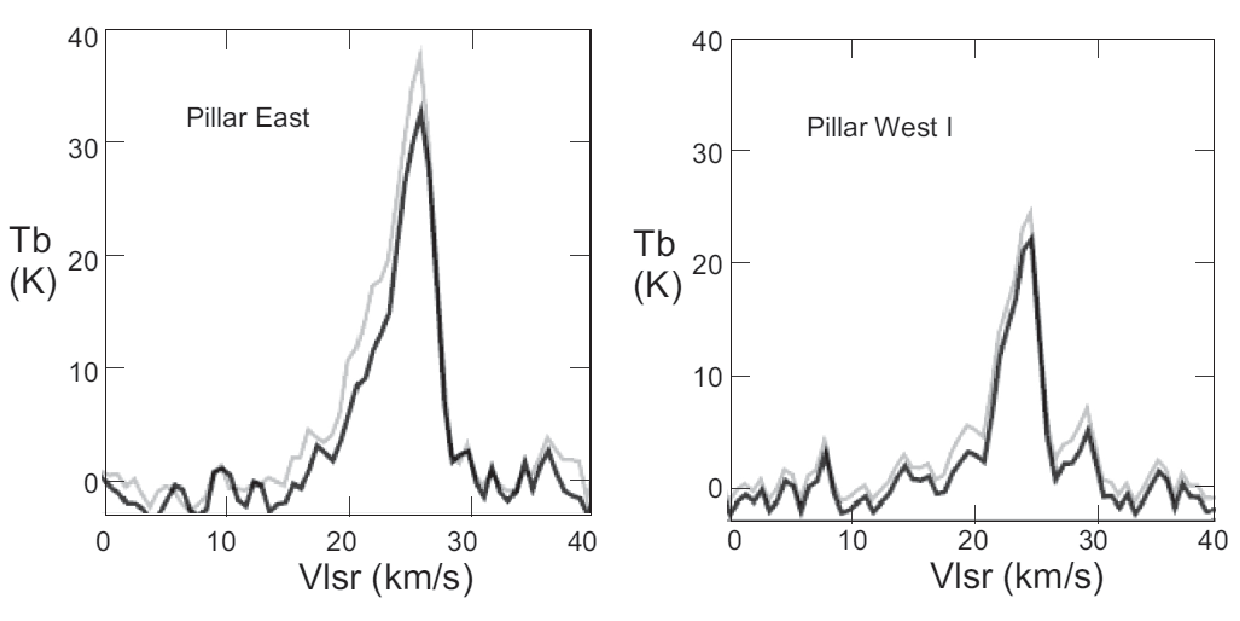}      
\end{center}
\caption{\co line profiles at the molecular tips of Pillar East and West I from \co line emission.   Full and grey lines show BGF and original spectra, respectively.}   
\label{co_line_profile} 
 	\end{figure*}

 Figure \ref{co_line_profile} shows the \co line profiles toward the tips of Pillar East and West I. In the analyses below, the diffuse back- and fore-ground emissions have been subtracted from the original maps using the background-filtering (BGF) technique (Sofue and Reich 1979) with a filtering beam of $4'.2$ (2.5 pc) in order to measure only the quantities belonging to the pillars.  

The line profiles show that Pillar West I has a peak brightness temperature $\Tb = 22$ K at radial velocity $\vlsr \sim 24.5$ \kms and the velocity full width is $\delta v \sim 3.5$ \kms (velocity dispersion of $\sigma_v\sim 1.8$ \kms), and $\Tb=33$ K at $\vlsr=25.5$ \kms with $\delta v\sim 3.6$ \kms for Pillar East.

 In the figure, original line profiles without BGF are also shown by grey lines, indicating that 5 to 10\% of of the tip emissions are due to contamination of the fore- or back-ground extended component. Note that the fractional contamination would be more significant in less intense regions along the pillars.

\begin{figure*} 
\begin{center}    
\includegraphics[width=17cm]{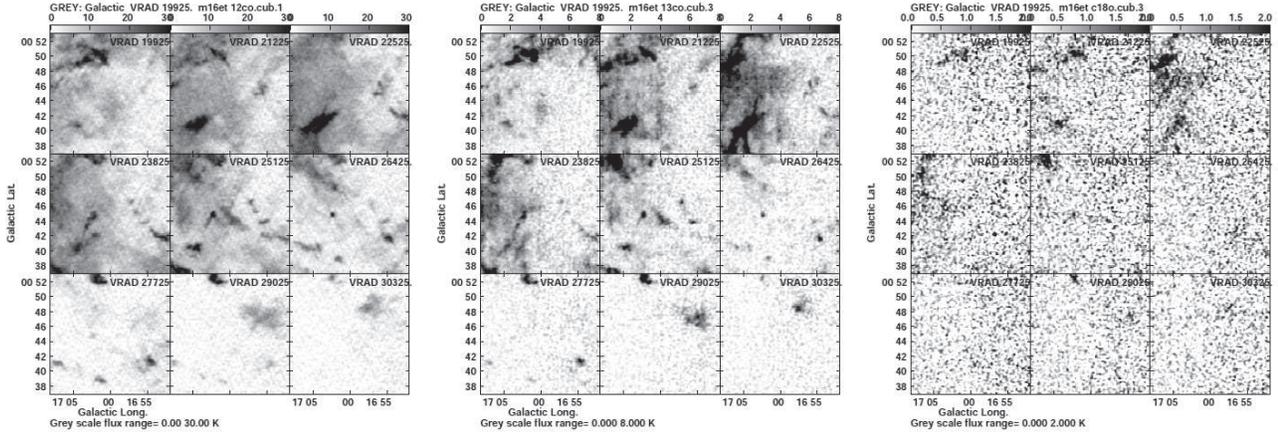}    
\end{center}
\caption{\co, \coth, \coei line channel maps (from right to left) of the M16 elephant trunks region as obtained by the FUGIN CO survey with the Nobeyama 45-m telescope. No BGF is applied. The M16 ETs, or the Pillars East and West, show up as two tilted ridges around the center at position angles $\sim 130\deg$ and $\sim 45\deg$, respectively, in channels at $\vlsr\sim 25$ \kms.  Grey scale unit is K of $\Tb$.} 
\label{co_ch} 
 	\end{figure*}

\begin{figure*} 
\begin{center}  
\includegraphics[width=17cm]{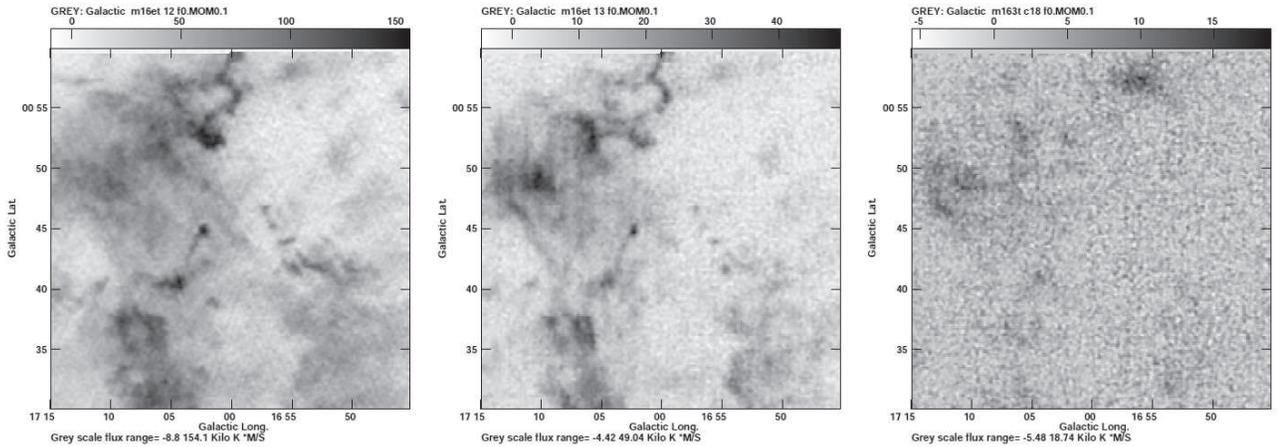}   
\end{center}
\caption{Integrated intensity maps (moment 0) of the \co, \coth and \coei lines of the M16 elephant trunks region as obtained by the FUGIN CO survey with the Nobeyama 45-m telescope. No BGF is applied.  Grey scale unit is \Kkms.} 
\label{co_m0} 
 	\end{figure*}

\begin{figure*} 
\begin{center} 
\includegraphics[width=15cm]{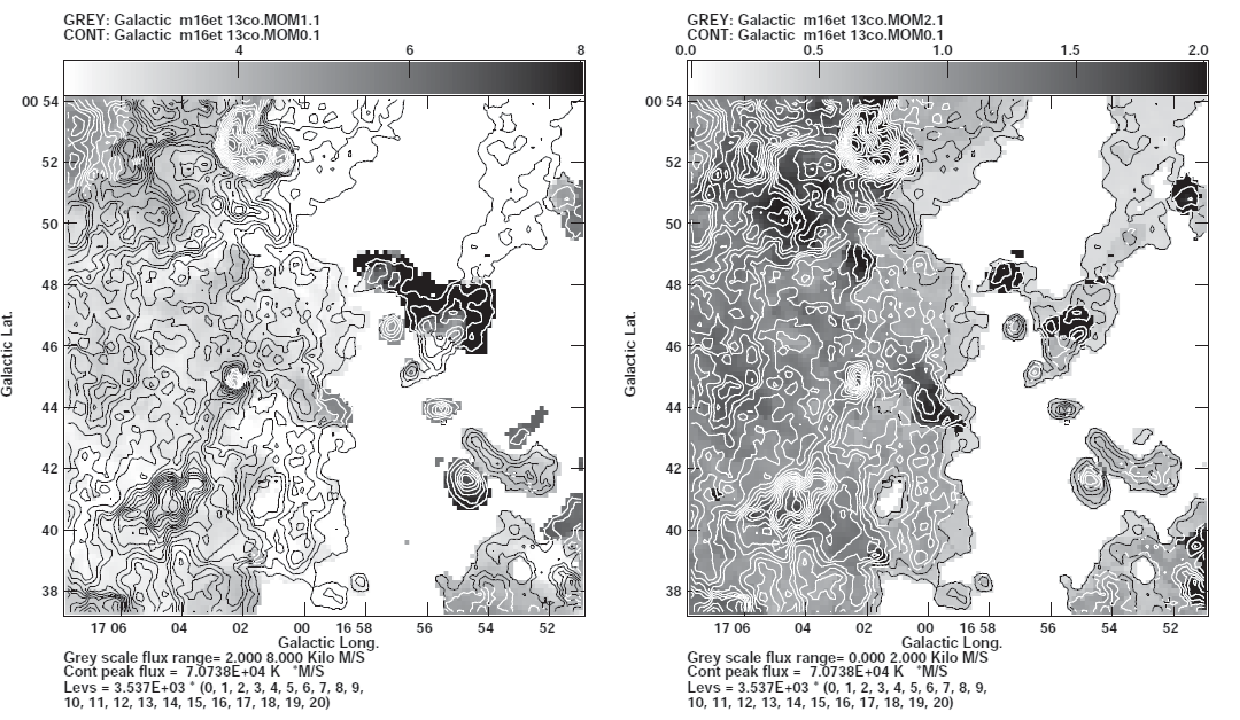}   
\end{center}
\caption{ Moment 1 (velocity field) and 2 (velocity dispersion) maps of the \coth overlaid with moment 0 contours for the central area of figure \ref{co_m0}.
 Grey scale unit in \kms.}  
\label{co_m12} 
 	\end{figure*}

Figures \ref{co_ch} and \ref{co_m0} show channel maps and integrated intensity maps between $\vlsr=20$ and 30 \kms of the M16 region including Pillar East, West I, II and III, where no BGF has been applied. The ETs are clearly recognized in the \co and \coth maps as the two obliquely extended ridges, approximately pointing the center of the OB cluster NGC 6611 at $(l,b)\sim(16\deg.95, 0\deg.8)$. The ETs show up from $\vlsr \sim 23$ to $\sim 27$ \kms, with the center velocity at 25 \kms. Figure \ref{co_m12} shows velocity field (moment 1) and velocity dispersion (moment2) in the same region overlaid with intensity contours.

\begin{figure*} 
\begin{center} 
 \includegraphics[width=10cm]{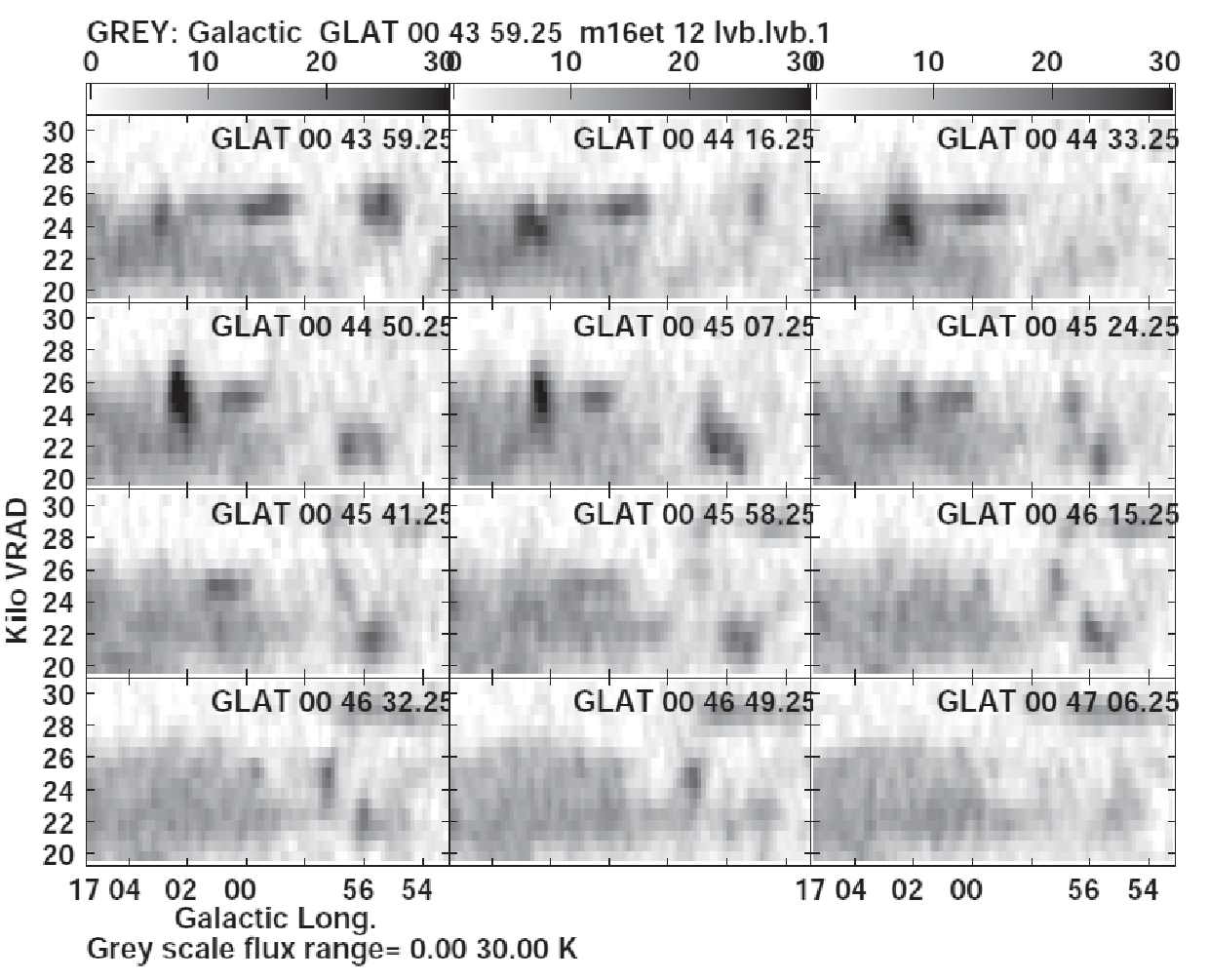}   
\end{center}
\caption{ Longitude-velocity (LV) diagrams of \co line emission at different latitudes across ETs East and West at $l\sim 17\deg 04''$ and $16\deg 56-58''$, respectively.   Grey scale unit is K in $\Tb$. }
\label{lv_ch}  
\end{figure*}

Figure \ref{lv_ch} shows longitude-velocity (LV) diagrams across Pillar East and West at different latitudes, where the ETs appear as intensity peaks at $l\sim 17\deg$ and $16\deg.95$, respectively. The LV ridges show slight inclination with respect to the vertical axis, showing rotation of the clump around its center. However, the directions of the inclination at different latitudes (channels) appear to be not uniform, but are rather random. Detailed discussion of the rotation of ETs will be given in the last section. 

\subsection{Radio continuum maps}

\begin{figure*} 
\begin{center}  
\includegraphics[width=14cm]{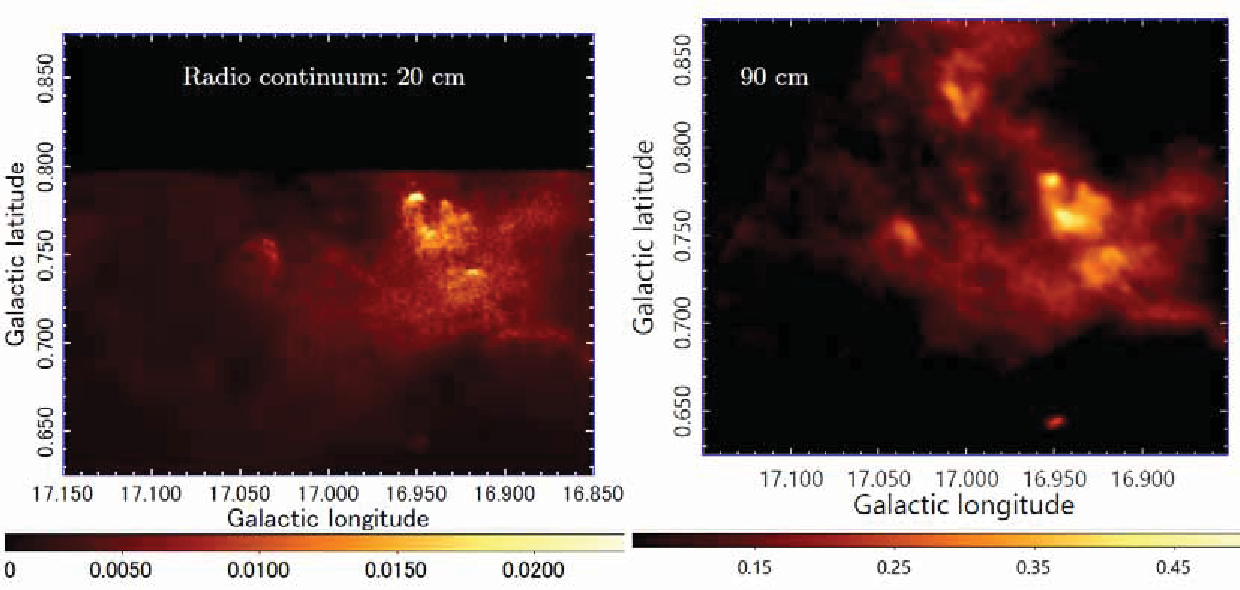}   
\end{center}
\caption{Radio continuum $\lambda$ 20-   and 90-cm maps of the M16 region including the pillars of creation in the same region as figure \ref{bb_8_co}. Note that the brightest radio source in the region coincides with the tip of Pillar I.  Color bars indicate radio brightness in Jy/beam.}   
\label{bb_20_90} 
 	\end{figure*}

 Figure \ref{bb_20_90} shows  radio continuum maps at $\lambda$ 20 and 90 cm extracted from the archival data from MAGPIS for the same region as in figure \ref{bb_8_co}. The 20 and 90 cm maps had angular resolution of $6''.2 \times 5''.4$ using B, C and D configurations, and $24''\times 18''$ using B and C configurations, respectively. Note that the upper 1/4 field is lacking in the 20-cm map.
Since the largest detectable angular scales were $16'$ and $70'$ at 20 and 90 cm, respectively, missing fluxes are negligible in the brightness measurements of the Pillars' tips.

The 20-cm emission exhibits  cometary structures in radio continuum, spatially coinciding with Pillar East and West I, II and III in FIR and CO line emissions. The 90-cm map also reveals the radio pillars as well as the extended thermal emission from the extended HII region of M16.   
Radio pillars compose bright head-tail structures coincident with the FIR and molecular ETs.

It is stressed that the radio source at Pillar I's tip is the brightest compact radio source in the M16 region, which is also coincident with the 8 \mum bright spot, although there appears slight displacement from each other.
Besides Pillar I to III, a broader, but fainter, cometary radio source is associated with Pillar East at $(l,b)\sim (17\deg.04, 0\deg.74)$.

\begin{figure*} 
\begin{center} 
\includegraphics[width=15cm]{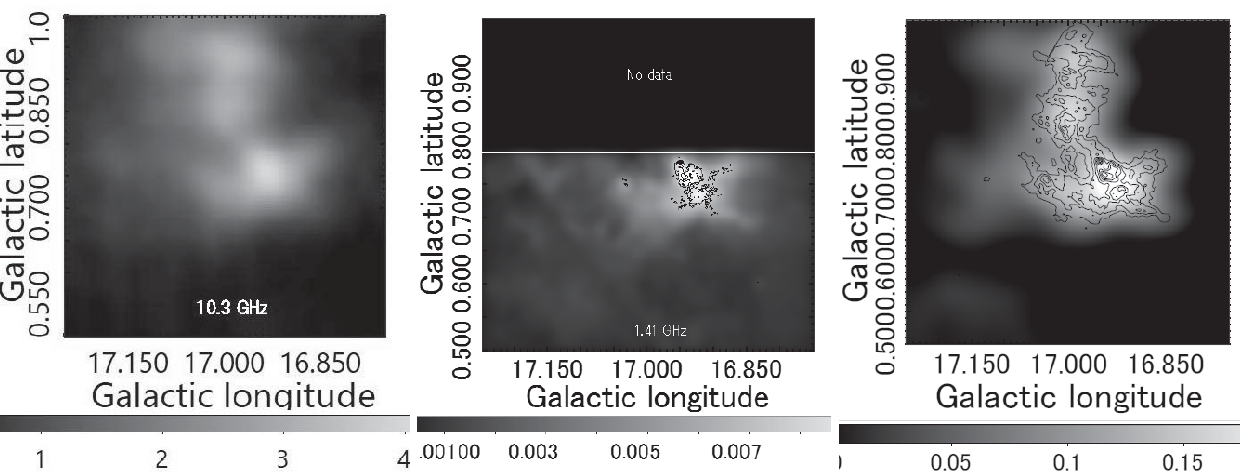}    
\end{center}
\caption{Radio continuum maps of the HII region M16 at 3, 20, and 90 cm smoothed to a beam of $2'.6$ (3 cm with 45 m telescope) in grey scales (Jy/original telescope beam by grey scale bars). Contours indicate the original resolution maps  only for references. } 
\label{r_sm} 
 	\end{figure*}

In figure \ref{r_sm} we show a radio continuum map at 3 cm (10.3 GHz) taken with the Nobeyama 45-m telescope (Handa et al. 1987) in comparison with the 20 and 90 cm maps smoothed to the same angular resolution of the 3-cm map of $2'.6$. Overlaid contours show the original resolution maps.

\section{Molecular Properties}
\subsection{Pillars in FIR, CO and Radio}
\begin{figure*} 
\begin{center}  
\includegraphics[width=14cm]{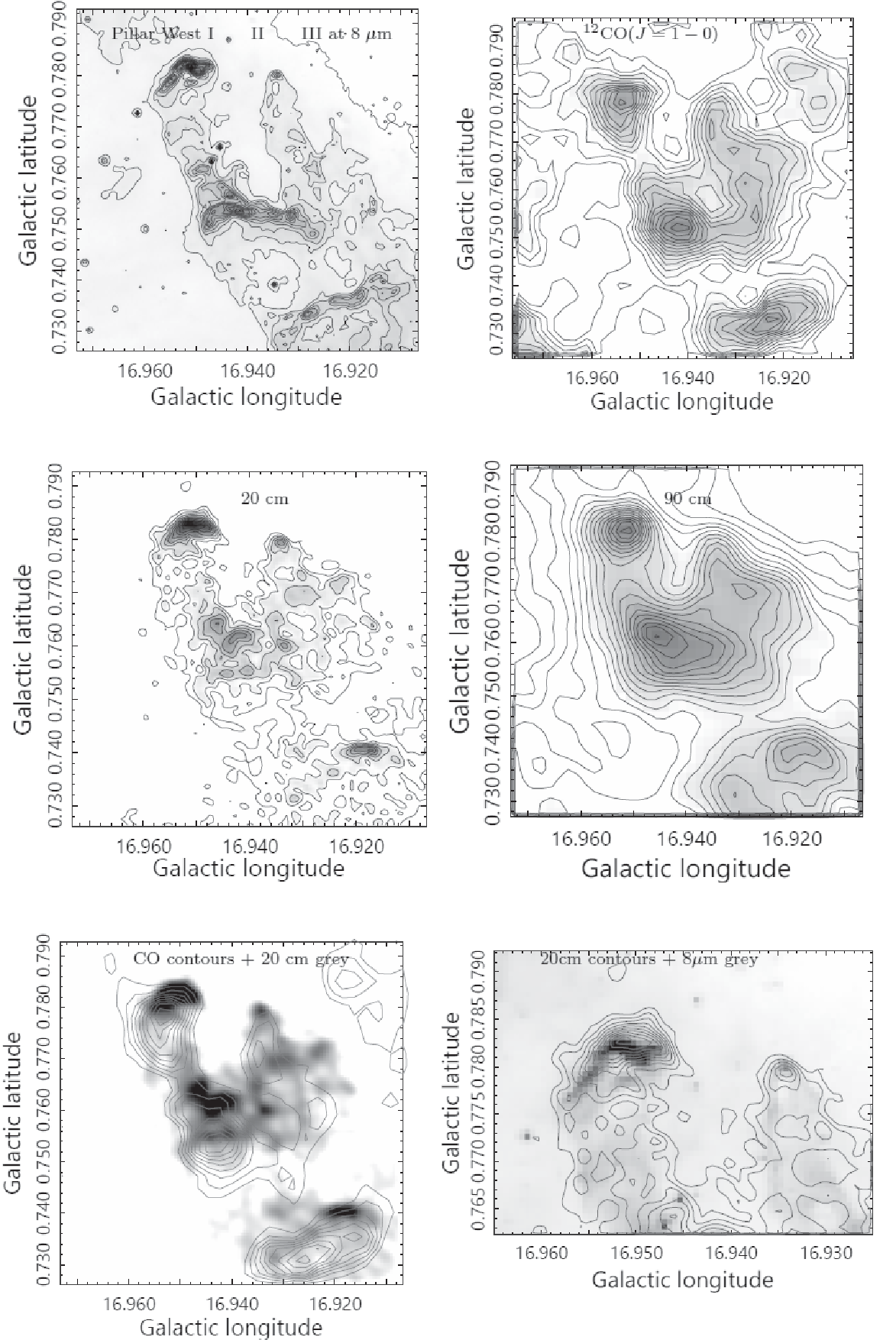}  
\end{center}
\caption{Pillars of Creation I, II and III at 8 \mum (contours from 50 at interval of 50 mJy/str), \co intensity (contours from 30 \Kkms at interval 5 \Kkms, radio continuum at 20 cm (contours from 0.008 Jy/beam at interval 0.02 Jy/beam), 90cm (from 0.1 by 0.02 Jy/beam, CO contours overlaid on 20 cm grey, and 20 cm contours on 8 \mum grey. The squared field is $(0\deg.07\times 0\deg.07)=(2.44{\rm pc}\times 2.44{\rm pc})$, except for the right-bottom panel $(0\deg.04 \times 0\deg.03)$. }
\label{pillars_all}     
 	\end{figure*}

In order to investigate the relation among the radio continuum, CO line, and FIR emissions in the pillars in further detail, we enlarge the pillars in figure  \ref{pillars_all}. The similarity of the 8 \mum and 20-cm radio maps is remarkable. The CO map shows significant displacement of the peak positions of the molecular gas column density from those traced by radio continuum emission. The 90 cm map has less resolution, while it agrees well with that at 20 cm. A detailed comparison of earlier CO and radio continuum maps with FIR images at various wavelengths has been obtained by White et al. (1999) and Urquhart et al. (2003).
 
\subsection{Molecular Pillar West I, II and III}

The \co intensity shows clumpy ridges along the three ETs with Pillar I being strongest. Using the line profile from the BGF data cube (figure \ref{co_line_profile}), we estimate the integrated intensity to be $\Ico = 77 \simeq \Tb \delta v$ \Kkms with $\Tb=22$ K and $\delta v=3.5$ \kms. Applying the CO-to-\Htwo conversion factor of $\Xco=2\times 10^{20}$ \Htwo/\Kkms (Bolatto et al. 2006), we obtain a column density of $\NHtwo=\Xco \Ico \simeq 1.54 \times 10^{22}$ \Htwosqcm.

\begin{table*}  
\caption{Molecular parameters of Pillar West I and East}
\begin{center} 
\begin{tabular}{llll} 
\hline 
\hline 
 & Pillar West I & Pillar East\\
\hline
Head \\
\hline
$(l,b)$ \dotfill & $(16\deg.96, 0\deg.78)$ & $(17\deg.04,0\deg.75)$ &(deg)\\
$\vlsr$ \dotfill & 24.5  	& 25.5 &(\kms) \\
Distance from Sun  \dotfill &2  	& 2 &(kpc)\\
Size parameter, $d=\sqrt{xy}$ \dotfill &0.27  &0.36 &(pc)\\
Full velocity width, $\delta v=2\sigma_v$  \dotfill & 3.5  & 3.6 &(\kms)\\
$\Tb ^\dagger$of \co \dotfill & 22 & 33& (K)\\
$\Tb $ of \coth \dotfill & 7 & 15 & (K)\\
$\Ico\simeq \Tb \delta v$(\co)\dotfill &77  &122 & (K \kms)  \\   
$N_{\rm H_2}$(\co, $\Xco$)  \dotfill &  $1.54\x 10^{22}$ & $2.3\times 10^{22}$&(\htwosqcm) \\
$N_{\rm H_2}$(\coth, LTE) \dotfill &  $1.61\x 10^{22}$ & $2.4\times 10^{22}$&(\htwosqcm) \\
$N_{\rm H_2}$(mean)  \dotfill &  $1.6\x 10^{22}$&  $2.4\times 10^{22}$ &(\htwosqcm)\\
$n_{\rm H_2}$ \dotfill &$1.9\x 10^{4}$  & $2.3\times 10^4$ &(\htwocc)\\
 $M_{\rm mol}^*$ \dotfill & 13.4 &39& ($\Msun$)\\ 
 $M_{\rm mol: photo}\ddagger$ \dotfill & $17.5\pm 0.5$ & $32.6 \pm 0.7$& ($\Msun$)\\ 
\hline 
\hline  
Tail \\
\hline
Length $X$  \dotfill &$\sim 6$ & $\sim 4$&(pc)\\
Width $Y$ \dotfill & $\sim 0.4$ & $\sim 0.4$& (pc)\\
${dv \/dX} \sin\ i$ \dotfill & $\sim \pm 1$  & 0.4&(\kms/pc)\\
${dv \/dY} \sin\ i$\dotfill & $\sim \pm 4$  & $-0.3$ & (\kms/pc)\\
Rotation $\sin\ i$ at edge\dotfill & $\sim \pm 0.4$ & $-0.06$& (\kms) \\
$\langle \Ico \rangle$ of \co \dotfill & $\sim 20$ &$\sim 30$ &(K \kms)\\
$\langle N_{\rm H_2}\rangle$ \dotfill & $\sim 4 \x 10^{21}$  &$\sim 6\times 10^{21}$ &(\htwosqcm) \\
$\langle n_{\rm H_2}\rangle$ \dotfill & $\sim 3\x 10^{3}$  & $3\times 10^3$ &(\htwocc)\\
$M_{\rm mol}$   \dotfill &$\sim 200$ & $\sim 210 $&$(\Msun)$\\  
\hline 
\end{tabular} 
\\
$^\dagger$ Diffuse components have been subtracted using the BGF technique.\\
$^\ddagger$ Photometric value enclosed by the circles in figure \ref{photometry}.\\
$^*$ The virial mass is much higher, indicating that the clump is not gravitationally bound (see subsection \ref{magconf}.)
\end{center}
\label{tab_pillars}  
\end{table*} 

The \Htwo column can be also calculated by assuming local thermal equilibrium (LTE) of CO molecules in the gas cloud using both the \co and \coth lines. The excitation temperature of the CO molecules is obtained by (Pineda et al. 2008)
\be
\Tex=5.53194\times  {\rm ln} \( 1+{5.53194\/\Tb(^{12}{\rm CO})+0.83632} \)^{-1} \nonumber \ee
\be
~~~~~~~~~~ \sim \Tb + 3 {\rm K}.
\ee
This can be used to estimate the optical depth of the \coth   line  as
\be
\tau(^{13}{\rm CO})=-{\rm ln} 
\(1-{\Tb(^{13}{\rm CO})/5.28864 \/1/(e^{5.28864/\Tex}-1)- 0.167667} \).
\ee 
The column density of $^{13}$CO molecules is given by 
\be
\Ncoth=2.4\times 10^{14} {\tau \Tex \/1-e^{-5.28864/\Tex}}\Delta v,
\ee
which yields the \Htwo column by
\be
\NHtwocoth=Y_{\rm ^{13}CO} \Ncoth,
\ee
where $Y_{\rm ^{13}CO}=7.7\times 10^5$ is the inverse of the abundance ratio of the $^{13}$CO molecules with respect to \Htwo  \ (Kohno et al. 2019). 

Inserting the observed peak brightness temperatures of \co and \coth lines at the clump head, $\Tb(12)=22$ K and $\Tb(13)=7$ K, respectively, and $\Delta v\sim 3.5$ \kms from the \co line profile, we obtain $\NHtwo=1.61 \times 10^{22}$ \Htwosqcm in agreement with the estimate from $\Xco$ conversion. The present values are consistent with the column density in the Pillars obtained by higher resolution interferometer observations by Pound et al. (1998). 

Taking the average of the column densities calculated from the \co intensity with $\Xco$ and \coth in LTE as $N_{\rm H_2}=(1.54+1.61)/2 \times 10^22 \simeq 1.6\times 10^{22}$ \Htwosqcm, the volume density in the head clump of Pillar I is estimated to be $\nhtwo=\Nhtwo/d \sim 1.9 \times 10^4$ \Htwo cm$^{-3}$, where $d=0.27$ pc is the diameter of the clump in the CO emission. 

\begin{figure} 
\begin{center}   
\includegraphics[width=6cm]{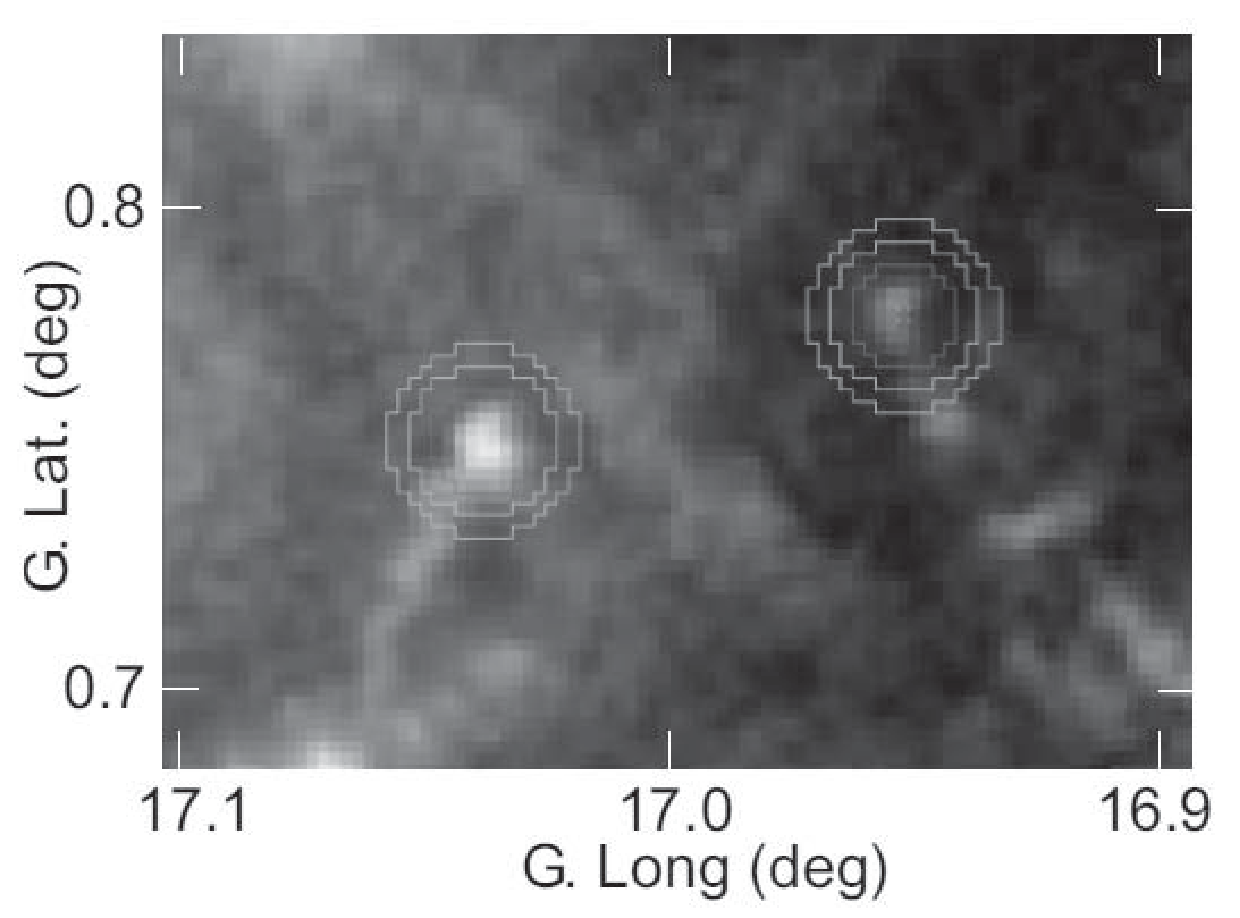}  
\end{center}
\caption{Photometry apertures in order to measure the total molecular masses of the head clumps of Pillar East and West I. The inner circles enclose the objects and outer two circles enclose the sky to be subtracted.} 
\label{photometry}
 	\end{figure}

The total molecular mass involved in a sphere of radius $d/2$ is then
$M_{\rm mol}\sim (4 \pi/3) (d/2)^3 \ 2.8 \mH \nhtwo \sim 13.4 \Msun.$
The molecular mass was measured also by integrating the observed intensities in the moment 0 map multiplied by the $\Xco$ factor, applying an aperture photometry tool to the circles shown in figure \ref{photometry}, where the inner circle enclose the sources and the outer two circles enclose the sky to be subtracted.  
Determined parameters and estimated quantities are summarized in table \ref{tab_pillars}, where the distance to M16 is assumed to be 2.0 kpc. 

 The estimated mass of Pillars West I may be compared with the CO($J=3-2$)-line observations by White et al. (1999), who report a molecular mass of $M_{\rm mol}\sim 60\Msun$ for Pillar I, and an even greater value, $M_{\rm mol}\sim 300 \Msun$, is reported by interferometer \co observations (Pound 1998).  
On the other hand, our result is comparable to those for ETs other than M16. CO observations toward ET clumps in Rosette Nebula indicate $n_{\rm H_2}\sim 10^4$ \htwocc and $M_{\rm mol}\sim 6-11 \Msun$ (Schneps et al. 1980), and ETs in four other HII regions of $\sim 6-29\Msun$ (Gahm et al. 2006).

In order to inspect into more detailed spatial and topological relation among the pillars in CO, radio continuum and 8\mum emissions, we show enlarged maps around the head clumps of the pillars in figure \ref{pillars_all}.

The head clumps in the radio continuum are significantly and coherently displaced from those in the \co line emission, in the sense that radio emission appears to originate at higher-latitude side surfaces of the CO clumps. This trend is seen also in the middle and bottom-side clumps on the pillars.

On the other hand, the radio head clumps are more closely correlated with the 8 \mum clumps, although that in Pillar I is also slightly displaced from 8\mum toward the higher-latitude direction, in the same sense as against CO.

\subsection{Molecular Pillar East}

\begin{figure} 
\begin{center}   
\includegraphics[width=7cm]{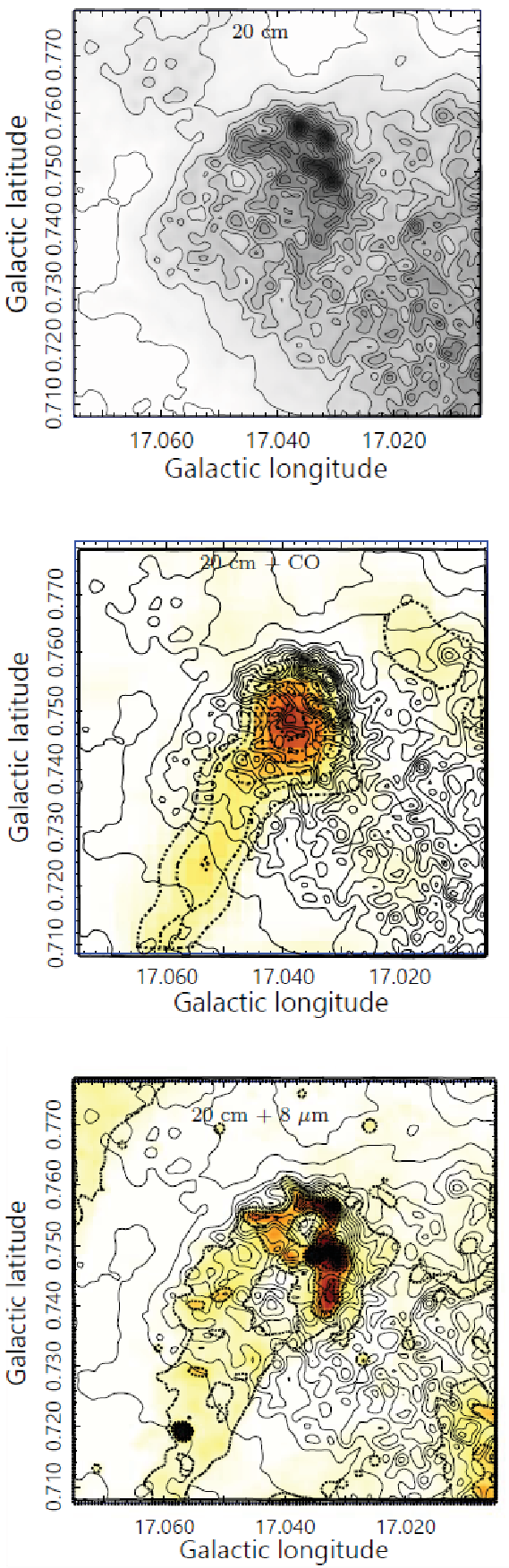}  
\end{center}
\caption{M16 ET East in radio continuum at 20 cm with contours from 0 at 0.01 Jy/beam interval, and overlays on \co intensity map in grey with dashed contours from 60 at 10 \Kkms interval, and on 8 \mum map with dashed contours from 100 at 10 Jy/str. The squared field is $(0\deg.07\times 0\deg.07)=(2.44{\rm pc}\times 2.44{\rm pc})$. }
\label{East_20_CO_8mu}  
 	\end{figure}

Pillar East is also well known for the HST images (Eagle's Pillar). Figure \ref{East_20_CO_8mu} shows an enlarged map of the 20-cm brightness around the tip as compared with the \co and 8 \mum maps. 
The 20 cm map shows a broad cometary structure with a well defined half shell surrounding the CO clump. The 8 \mum emission shows also shell feature nearly coincident with the radio shell. The hole in 8 \mum coincide with the CO intensity peak. Thus, the molecular pillar composes the back bone of the structure,  which is enveloped by the FIR, also optical, and radio continuum sheathes. The estimated parameters are listed in table \ref{tab_pillars}.
 
\section{Radio continuum properties}

\subsection{Cometary radio cap}
The 20 cm continuum map shows that the radio sources at the tips of Pillars East and West I have a cometary cap structure. They are significantly off-set from the molecular head clumps, and are concave to the center of molecular gas distribution. The radio caps spatially coincide with the 8\mum bright rims with slight off-set outward. 
The radio + CO structure at the Pillars' heads shows a close resemblance to that observed in bright-rimmed molecular globules associated with conical radio continuum rims produced by photo-ionization by nearby O stars (Urquhart et al. 2006). Also, our result will prove the current theoretical models including the radiation-driven implosion mechanisms ({Haworth \& Harries} {2012}; and the literature therein).

\subsection{Thermal radio spectrum}
\begin{figure} 
\begin{center}  
\includegraphics[width=7cm]{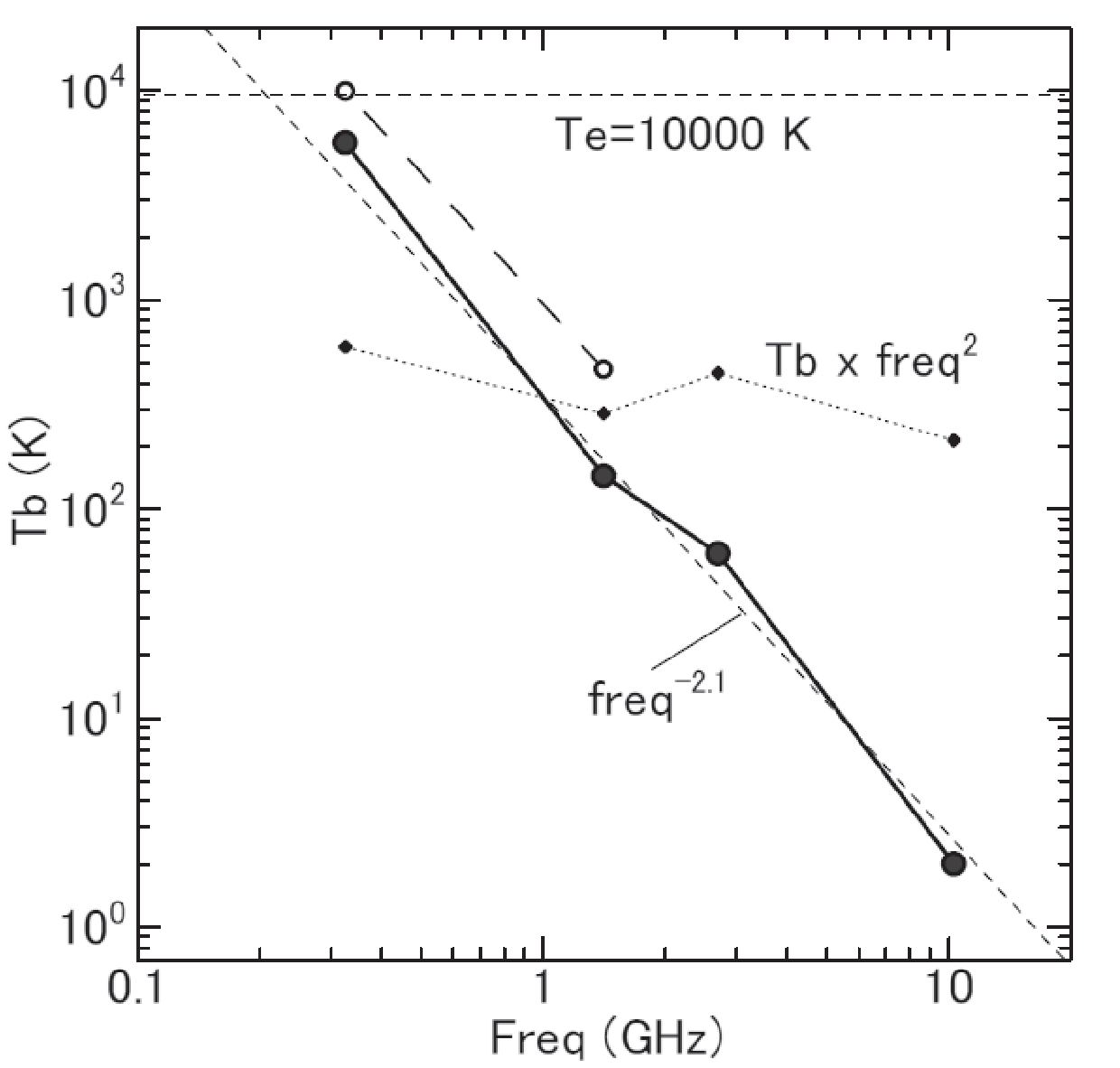}  
\end{center}
\caption{Radio continuum spectrum of Pillar I's peak. Plotted are the peak surface brightness temperatures, $\Tb$, in K at 90, 20, 11 and 3cm smoothed to 3-cm beam ($2'.6$) except for 11 cm with original resolution.  } 
\label{spectrum} 
 	\end{figure}

Using the smoothed maps in figure \ref{r_sm}, we plot a radio continuum spectrum between 10.3 GHz (3 cm) and 0.33 GHz (90 cm) of the head clump of Pillar I in figure \ref{spectrum}. We also plotted the brightness at 2.7 GHz (11 cm) in original resolution of $4'$ from the Bonn 100-m telescope (Reich et al. 1984, 1990). Small open circles show the brightness in the original resolution at 20 and 90 cm.  

The radio spectrum is well fitted by that expected for  optically-thin  thermal (free-free) emission with a spectral index of $\beta=-2.1$ ($\Tb \propto \nu^\beta$) from ionized gas of an HII region. Since the head clump is nearly resolved in the present VLA maps, the brightness temperature at 90 cm at original resolution, $\Tb \sim 10^4$ K, represents the lower limit to the electron temperature of the brightest spot, which indicates that $\Te \ge 10^4$ K, significantly higher than that observed in extended HII regions, which are observed to have $\Te \sim 7000-8000$ K (Downes et al. 1980). Accordingly, the turn over frequency would be lower than 0.3 GHz, because the spectrum is nearly straight between 0.33 and 10 GHz.

\subsection{Radio Pillar West I}
We now derive radio continuum parameters for the  Pillar  I's head cap.
The 90-cm map shows that the peak brightness temperature is as high as $\Tb\simeq 10^4$ K,  indicating that the electron temperature is equal to or higher than this value, $\Te \ge 10^4$ K.  
In the following, we estimate the density of ionized gas assuming $\Te=10^4$ K. 

The brightness temperature at 20 cm at the peak is measured to be $\Tb=484$ K, which is sufficiently smaller than the electron temperature, so that we may assume that the optical depth at 20 cm is small,
$
\tau=\Tb/\Te \sim 0.048 \ll 1.
$
Inserting this into the relation with the emission measure $EM$ and frequency,
\be
\tau=0.082 (\Te/K)^{-1.35} (\nu/{\rm GHz})^{-2.1}EM,
\ee
we obtain the emission measure to be
$EM=3.05\times 10^5$ pc cm$^{-6}$.

From the 20-cm radio map, the size diameter, $d$, of the head clump is measured to be
$
 d = \sqrt{xy}\simeq 0.16
$
pc, where $x=0.23$ and $y=0.12$ pc are the major and minor axial full widths of the emission region. Assuming that the line-of-sight depth of the radio clump is equal to $d$, we obtain the electron density to be
$
\nel\simeq 1.36\times 10^3$ cm$^{-3}.
$
For these density and size, the mass of the ionized gas is on the order of $\sim 0.1 \Msun$.
 The electron density is consistent with that obtained by optical spectroscopy by McLeod et al. (2015), who derived maps of electron density and temperature, showing $\nel \sim 1-2\times 10^3$ cm$^{-3}$ and  $\Te \sim 1-2\times 10^4$ K in the pillar tips.  

Thus, the radio continuum clump at Pillar I's head can be categorized as a compact HII region of relatively small size. Pillar I in radio exhibits approximately the same size as, similar cometary morphology to, an order of magnitude less EM than, and several times less electron density than those observed in typical compact HII regions ({Deharveng \& Maucherat} {1978}; {Woodward, Helfer, \& Pipher} {1985}).

Given the electron density, we may estimate the luminosity of the recombination emission from the clump as
$L_{\rm reco}\sim \alpha \nel^2 V E_{\rm UV}\sim 487 \Lsun$,
where $\alpha \sim 4\times 10^{-13}$ cm$^3$ s$^{-1}$ is the recombination coefficient at $\Te\sim 10^4$ K, $E_{\rm UV}=13.6$ eV is the ionization energy of a neutral hydrogen, and $V=4 \pi/3 (d/2)^3$ is the volume of the HII region. The radio luminosity due to the free-free emission is negligible.
The estimated parameters are listed in table \ref{tab_radio}.

\begin{table}  
\caption{Radio continuum property of the head clump of Pillar West I and East}
\begin{center} 
\begin{tabular}{llll } 
\hline 
\hline 
Parameter   & Pillar West I& Pillar East \\  
\hline
$\nu$ \dotfill &1.41 &1.41& (GHz) \\
$\Tb$ \dotfill & 484 & 172 &(K)\\
$T_{\rm e}$ (assumed) \dotfill & $10^4$ & $10^4$& (K)\\
$\tau$ \dotfill & 0.048 & 0.017\\
$EM$ \dotfill & $3.05 \times 10^5$  &$1.1 \times 10^5$&(pc cm$^{-6}$)\\
$d=\sqrt{xy}$ \dotfill & 0.16 &0.18&(pc)\\
$n_{\rm e}$ \dotfill & $1.36 \times 10^3$ &$0.8 \times 10^3$ &(cm$^{-3}$ )\\  
$M_{\rm HII}$\dotfill & 0.07 & 0.06 &($\Msun$)\\
$L_{\rm reco}$ \dotfill & 487 & 244 &($\Lsun$)  \\
Dist. to O star & 2.0 & 3.2 & (pc)\\
$L_{\rm source}$ & $6\times 10^5$ &$5\times 10^5$ &($\Lsun$)\\
\hline
\end{tabular}  
\label{tab_radio}
\end{center}  
\end{table}  

The recombination time of the clump is therefore as short as
$t_{\rm reco}\sim \nel k \Te V/L_{\rm reco}  \sim 2$ y, where $k$ is the Boltzman constant. Note, however, that this time scale is a lower limit, because the electron temperature could be higher, as was expected from the high brightness temperature at 90 cm.
Equivalently, the cooling time of partially ionized hydrogen gas can be estimated using the cooling rate of $\Lambda\sim 10^{-23}$ erg cm$^3$ s$^{-1}$  at $\Te\sim 10^4$ K (Foster et al. 2012), which yileds $t_{\rm cool}\sim \nel k \Te /(ne^2 \Lambda)\sim 3$ y.
Thus, the gas must be heated either from inside or from outside in order to keep its luminosity, if the radio continuum clump is a stable structure.


One of the possible mechanisms of heating of Pillar I tip is photo-ionization by UV radiation from stars with effective temperature higher than $\sim 10^4$ K embedded in the HII region. However, there has been no report of such a star or a cluster of luminosity comparable to $\sim 487 \Lsun$ in the direction of the Pillar I's head (Xu et al. 2019), although there remains a possibility that such stars suffer from heavy extinction and are not recorded in the optical catalog of Evans et al. (2005), based on which the stellar distribution has been studied. In fact it is reported that a young stellar object equivalent to a B type star of several solar masses with $\sim 200 \Lsun$ exists embedded in the molecular tip of Pillar I (McCaughrean \& Andersen (2002); {Thompson et al}{2002}), while it is unclear whether the star is luminous enough to maintain the compact HII region.

Another possible mechanism is the excitation by UV photons from NGC 6611 about $D\sim 2$ pc away to the galactic north. Then, the required luminosity of the exciting stars is estimated to be
$L_{\rm OB}\sim L_{\rm reco}(4 \pi D^2/(\pi (x/2)^2))\sim 6 \times 10^5 \Lsun$, which is consistent with the luminosity of two O5 stars in the center of NGC 6611 (Pound et al. 1998).
  
\subsection{Radio Pillar East}

In the same way, the radio property of Pillar East can be estimated as follows: The brightness temperature is measured to be $\Tb\sim 172$ K, which yields an emission measure of $EM\sim 1.2 \times 10^5$ pc cm$^{-3}$. The line of sight depth is estimated to be $d\sim 0.18$ pc.  Then, we obtain an electron density of $\nel \sim 0.8 \times 10^3$ cm$^{-3}$. The recombination luminosity is $\sim 244 \Lsun$, and the required luminosity of the exciting stars at 3.2 pc away is estimated to be $L_{\rm source}\sim 5\times 10^5 \Lsun$, consistent with the estimation from the luminosity of Pillar I.

Because of the well resolved structure in the radio map of Pillar East, we may investigate the intensity distribution in more detail than for Pillar  I, which will be useful to clarify the relation to the O stars in NGC 6611 that are supposed to be the UV photon source.

\begin{figure} 
\begin{center}   
\includegraphics[width=7cm]{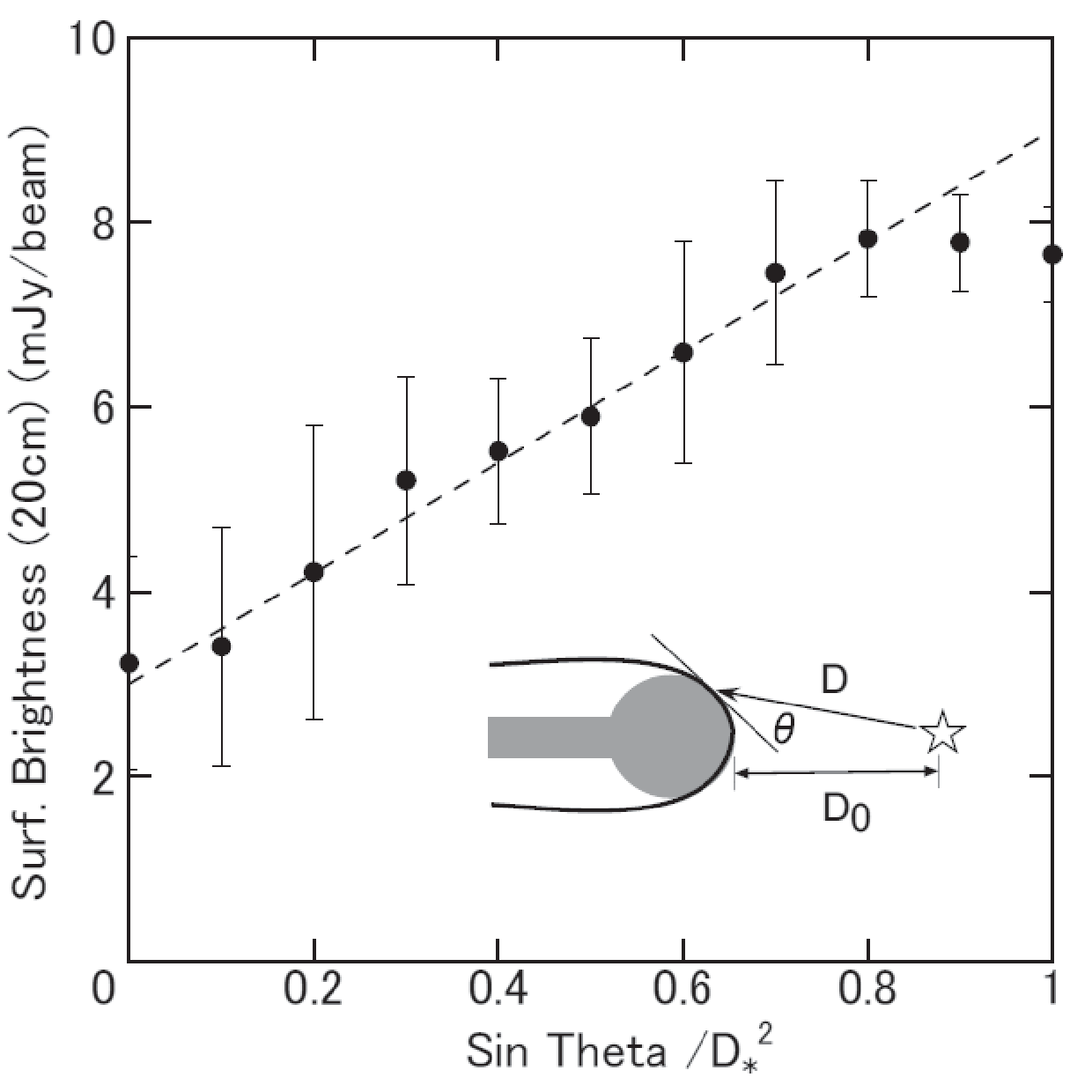}  
\end{center}
\caption{Radio intensity at 20 cm in mJy/beam along the surface of the cometary HII region of Pillar East against the injection-angle parameter, $\sin \theta/ D_*^2$, where $D_*=D/D_0$ with $D_0$ being the distance of the tip from the center of NGC 6611. The linear correlation is remarkable. Saturation at $\theta=90\deg$ is due to the finite resolution as well as to intrinsic thickness. The finite value at $\theta\sim 0\deg$ is due to the Galactic background.  } 
\label{angleSigma} 
 	\end{figure}

Figure \ref{angleSigma} shows radio intensity at 20 cm along the surface edge of Pillar East plotted against calculated injection-angle parameter defined by
$\chi=\sin \theta (D/D_0)^{-2}$. Here, $\theta$ is the angle of the pillar's surface as seen from the supposed center of UV radiation at the O5V star in NGC 6611, and $D$ and $D_0$ are the distances of the measured point and the tip of pillar from the star, respectively. Each point indicates Gaussian-weighted running mean of measured intensities at every 0.1 interval of $\chi$ with a half width of 0.15. The bars are standard deviations.

The figure shows a linear relation   on $\sin \ \theta$ as expected for illumination from outside. Note that the finite intensity at $\theta \sim 0$ is due to the background emission of the Galaxy. The saturation at $\theta \sim 90\deg$ is due to smearing both by the finite angular resolution and intrinsic thickness of the radio shell. 

\subsection{Radio brightness and distance from exciting star}
Besides the dependence on the injection-angle parameter within a pillar, variation of the peak brightness in different pillars with different distances from the exciting star would be another mean to clarify if the pillars have a common excitation source. 

\begin{figure*} 
\begin{center}  
\includegraphics[width=14cm]{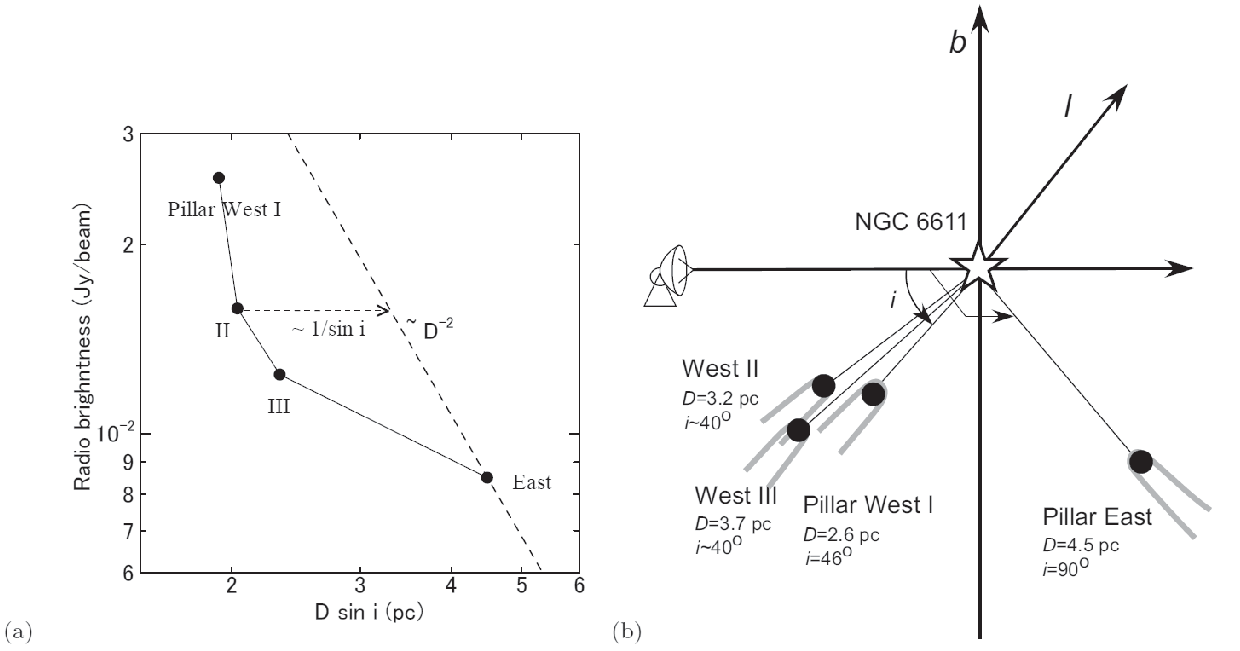}    
\end{center}
\caption{(a) Radio continuum peak brightness at 20 cm of the ET (pillar) heads as a function of the projected distance $D\ \sin \ i$ from the brightest O star of NGC 6611. Inserted by dotted line is the relation $\propto D^{-2}$ in arbitrary scale. (b) 3D positions of the pillars when Pillar East is assumed to have $i=90^\circ$.  } 
\label{radio_distance} 
 	\end{figure*}

In figure \ref{radio_distance} we plot the 20-cm peak surface brightness against projected distance, $D\ |\sin \ i|$, of the head clumps of the four pillars discussed in this paper from the brightest star of type O5V of NGC 6611. Here, $i$ is the inclination angle of the star-pillar tip line with respect to the line of sight (los) with $i=0^\circ$ indicating the object on the near side los. 
 The plot shows that the surface brightness decreases with the projected distance in agreement with the 3D orientation derived from optical spectroscopy and extinction analyses (McLeod et al. 2012). 

However, the slope of the plot is shallower compared with the square-inverse law about distance. Such displacement may be attributed either to variable inclinations or to different excitation parameters among the tips. However, the latter may not be the case, because the pillars are formed in the same HII region and their excitation and emission mechanisms would not be so different from a tip to the other. 

Assuming that the displacement is due to variable inclination, we may derive the 3D orientation of the pillar tips by adjusting $i$, so that the four tips fall on the $D^{-2}$ line. Because $|\sin \ i|\ \le 1$, Pillar East may be put farthest from the excitation center, and we here assume $i=90^\circ$ (on the plane of the sky). 
Accordingly, the 'true' distances of the four tips from the exciting star are assumed to be $\sim 2.6$, 3.2, 3.6, and 4.5 pc for West I, II, III and East, respectively. 
Then, Pillars West I, II and III are located at inclinations $i\sim 47^\circ, \ 40^\circ,$ and $40^\circ$, respectively, while mirror positions with respect to the sky plane are not excluded. 
Figure \ref{radio_distance}(b) illustrates the locations of the pillars with respect to the exciting star at the origin of the coordinates.

We emphasize that the linear relation against the injection-angle parameter in figures \ref{angleSigma} and the dependence on the distance in \ref{radio_distance} prove that the heating of the pillars are commonly driven by the radiation from the central O stars in NGC 6611. 

  \section{Discussion} 

\subsection{Implosion by HII pressure}

\begin{figure} 
\begin{center}  
\includegraphics[width=7cm]{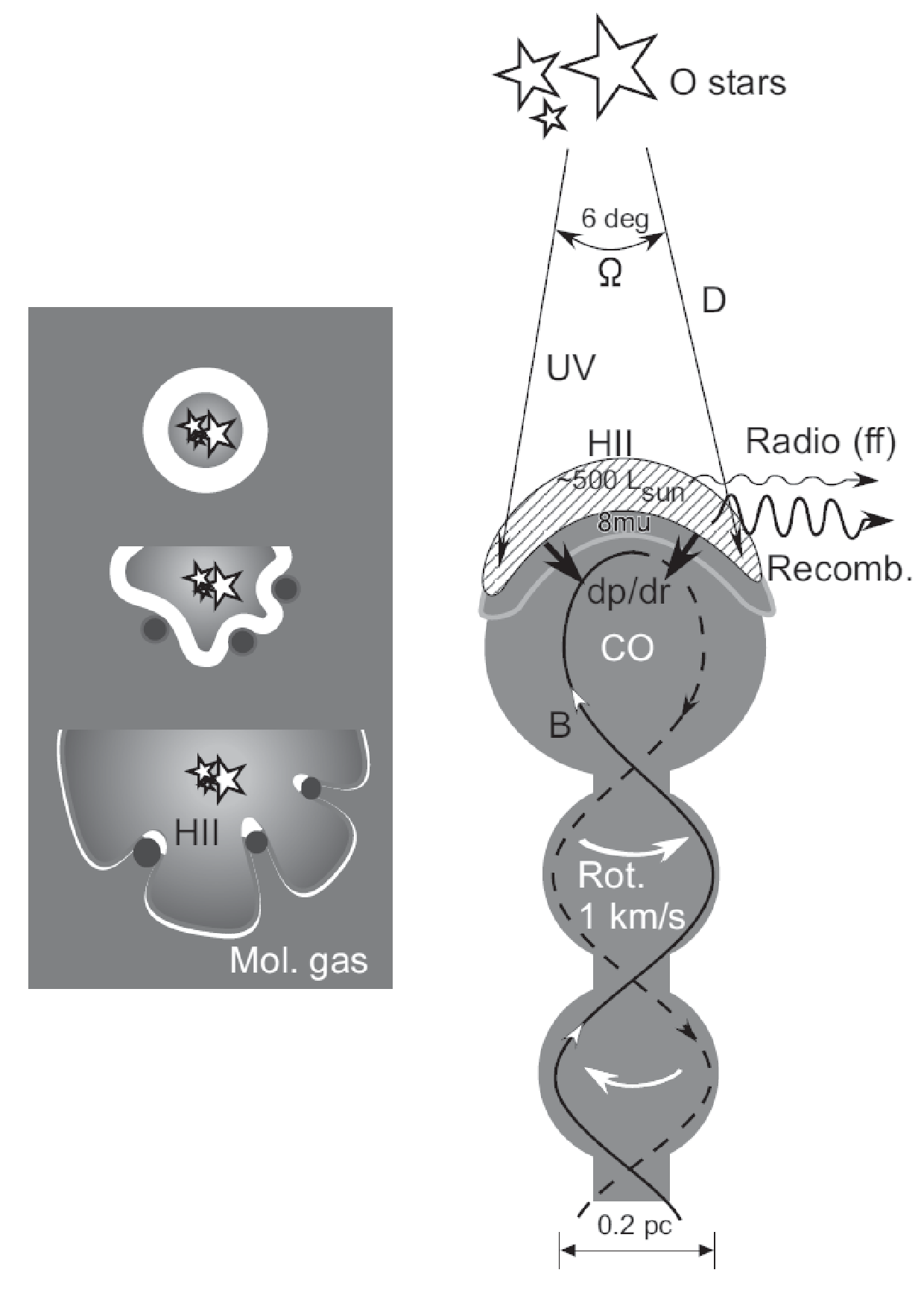}  
\end{center}
\caption{Schematic evolution of elephant trunks in radio continuum, and spatial relation between NGC 6611 and Pillar I. UV radiation from the two O5 stars of NGC 6611 of $L_{\rm UV}\sim 7 \times 10^5 \Lsun$ in the solid angle of Pillar I head is comparable to the recombination luminosity of the head clump. Torsional magnetic field will drive rotational oscillation of the clumps. } 
\label{illust}   
\end{figure}

Figure \ref{illust} schematically summarizes the spatial relationship among the emission features at the head clump of Pillar I in radio continuum, FIR, and CO line emissions. From such topology, we may draw a scenario to explain the astrophysical processes around Pillar I's head.  In the following orders-of-magnitude estimates, we use the projected distances, instead of the derived distances in the previous subsection, allowing for uncertainties of a factor of $\sim 2$. 

The entire M16 HII region is widely extended and the central part has already been exhausted to compose a cavity around the OB cluster NGC 6611. The Pillars are the closest relic of high-density molecular clump, where UV photons from the O stars are heavily shadowed, so that the pillars' tails are protected against dissociation. The strongly illuminated surface of the Pillar's edge is ionized, and generates compact HII region with high density and electron temperature, which shines as the brightest radio source in M16.

The compact HII region at the Pillar I tip is concave with respect to the molecular clump, so that the radiation as well as the ionized gas pressures accelerate the molecular surface toward the center of the clump, leading to implosive compression. 

The radiation pressure of UV photons from the O stars of NGC 6611 is on the order of 
$P_{\rm OB} \sim L_{\rm OB}/(4 \pi D^2)/c \sim 1.7\times 10^{-10}$ erg cm$^{-3}$. 
This pressure is equivalent to the pressure caused by the expelled ions from the surface of the neutral gas surface. 
On the other hand, the gaseous pressure of the HII gas in the radio cap, when it is maintained there stationary, is  
$P_{\rm HII} \sim \nel k \Te \sim 1.9 \times 10^{-9}$ erg cm$^{-3}$,
an order of magnitude more effective than radiation or the expelling momentum transfer in order to compress the gas cloud.

These pressures are  compared with the thermal pressure and turbulent energy density of the molecular clump on the order of
$P_{\rm mol}\sim 2 \nhtwo k \Tex \sim 1.4 \times 10^{-10}$ and 
$P_{\rm turb}\sim 1/2 \rho \sigma_v^2 \sim 1.5 \times 10^{-9}$ erg cm$^{-3}$,
where $\rho=2.8 \mH \nhtwo$ is the gas density and $\sigma_v\sim 1.8$ \kms is the velocity dispersion of the gas, which is half the full velocity width.

 It is thus found that the external pressure by the HII gas pressure is comparable or slightly higher than the internal pressure of the head clump. Assisted by the concave structure with respect to the cloud center, it would enhance the implosive compression.
We point out that the HII pressure is more effective than the radiation pressure from OB stars for implosion, which would give an observational constraint on triggering mechanism for star formation in ETs heads ({Haworth \& Harries} {2012}; and the literature therein). 

We here encounter a momentum problem. The total HII mass is only $\sim 0.1\Msun$, much less than the clump's molecular mass of $\sim 13\Msun$. 
 Therefore, the HII cap alone cannot give sufficient momentum to the cloud for contraction, because the HII gas will be evaporate within a crossing time of the sound velocity through the cap, $\sim 10^{4}$ y. In order for the HII cap to act to compress the molecular tip, it must be confined to the pillar by external pressure of the extended HII gas in M16, while the radiation pressure is not enough as discussed above. 
 
 We estimate the emission measure from the brightness of $\sim 0.005$ Jy/beam at 20 cm near the mid-point between the pillar and NGC 6611 to be on the order of $EM\sim 5\times 10^3$ pc cm$^{-3}$. If the line of sight depth is on the order of $\sim 10$ pc through M16, we obtain $\nel\sim 20$ cm$^{3}$. Then, the electron temperature of the extended HII must be higher than $7\times 10^5$ K, an X-ray temperature, in order to confine the HII cap to the pillar tip, which is, however, not a realistic value. 

Therefore, we need another mechanism to confine the HII gas near the molecular clump in order for the DRI works to dynamically compress the molecular clump. A possible mechanism to confine the HII gas would be the strong magnetic field, as will be discussed in the later section.
 
\subsection{Evolution of pillars}

Radio continuum maps revealed open cone morphology of the ionized gas composed of a bright cap of high density HII gas covering the tip and a cylindrical tail fading away along the molecular pillar.
 Although such open cometary structure is often observed in compact HII regions, where a wind from an O star or a gas flow associated with an expanding gas are the origin (Reid and Ho 1985), the present pillars appear to be not the case: 
 The dynamical pressure due to a wind from the O type, whose mass loss rate is on the order of ${\dot M}\sim 10^{-5}$ y$^{-1}$ ({Markova et al.} {2004}), is several orders of magnitudes less than the required dynamical pressure to affect the HII region of density $\sim 10^3$ cm$^{-1}$ at the pillar tip. Similarly, gas flow of the expanding HII region around NGC 6611, whose density is on the order of $\sim 10$ cm$^{-3}$ and expansion velocity of $\sim 10$ \kms, is also too weak to affect the pillar's dynamics.

 Along the lines of the currently accepted scenario for the formation of pillars in HII regions often referred to as radiation-driven implosion (RDI) mechanisms (Schneps, et al {1980}; 
Bertoldi 1989; 
{Whalen \& Norman} {2008};  
Gritschneder et al. 2010; {Mackey \& Lim} {2010}: 
Haworth \& Harries {2012}),
we here consider the formation and evolution of the M16 Pillars as the following (figure \ref{illust}).  

When the OB cluster, NGC 6611, was born, the entire M16 region was deeply embedded in a dense molecular cloud. As the HII sphere expands, the molecular wall is pushed to cause Rayleigh-Taylor instability and radiation driven implosion grow to form wavy surface with dense cores left behind as the heads of elephant trunks. The inner surface of the molecular cavity is ionized, and high-density HII shell is produced, which also follows the wavy growth of instability.  

As the waves grow, retarded portions of the front evolve into stretched elephant trunks. Accordingly, the UV flux per unit area on the molecular surface get highly non-uniform due to the variation of the injection angle, $\theta$, as well as the distance, $D$, from the excitation stars. 

As the ET grows, the angle on the side wall decreases to $\theta \sim 0\deg$ or even negative, and the photo-dissociation of the surface ceases. On the other hand, the top of the head clump, which is much closer to the stars, is still strongly illuminated at $\theta\sim 0\deg$, when the side wall is shadowed. Such scenario is partly proved by the fact that the radio brightness along the surface of the pillar obeys the simple illumination law as shown in figure \ref{angleSigma}.

\subsection{Rotation }

\begin{figure} 
\begin{center}  
\includegraphics[width=7cm]{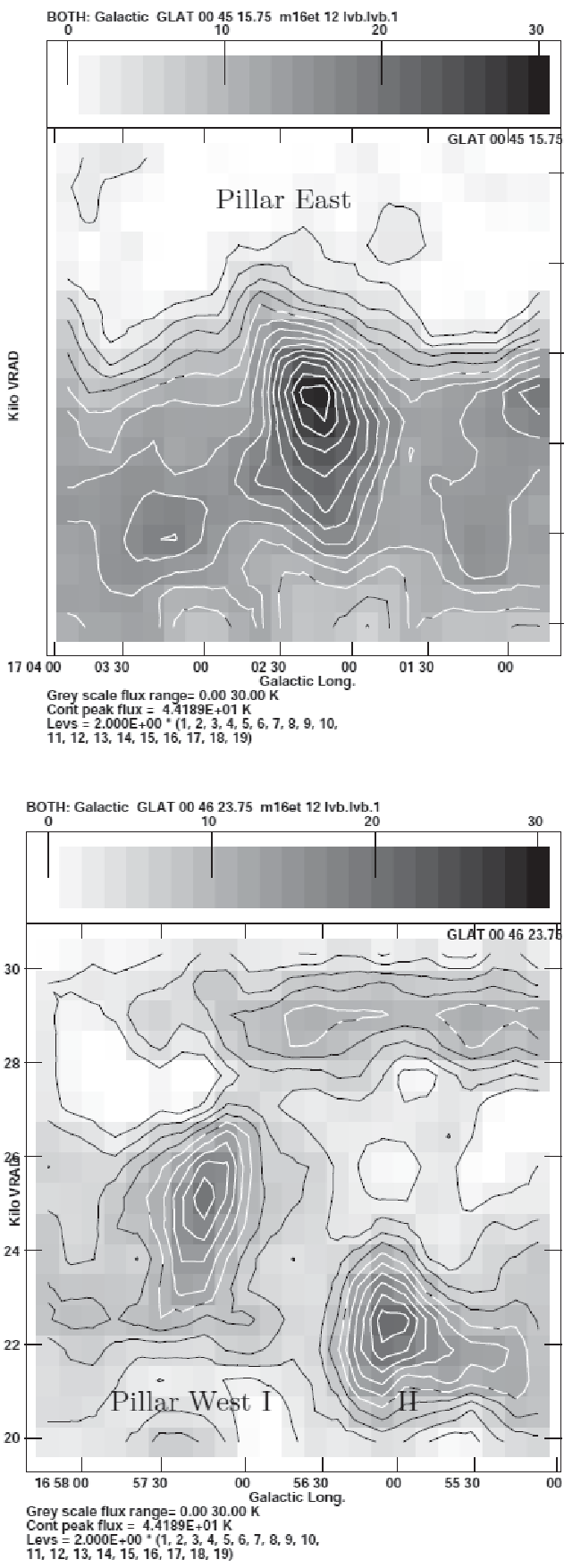}  
\end{center}
\caption{LV diagrams across Pillar East at $b=0\deg.754$ and Pillar  I and II at $b=0\deg.773$, showing the rotation of the trunk tails.}
\label{LV}  
\end{figure}

Figure \ref{lv_ch} showed tilts of intensity ridges on the LV diagrams. Such velocity gradients are more clearly seen in the clumps on the tails than in the heads. Figure  \ref{LV} shows LV diagrams of the tails of Pillar East at $b\sim 0\deg.75$ and Pillar I and II at $b=0\deg.773$. 
The velocity gradient  is $dv/dl \sim + 2{\rm km \ s^{-1}}/20''\sim 1.0$ \kms/0.1 pc across the tail of Pillar East, and $\sim -2{\rm km \ s^{-1}}/15''\sim 1.4$ \kms/0.1 pc across Pillar I and II. 

 The velocity gradient may represent either shear motion in an unbound clump, or rotation (spin). In the former case, the pillar would be destroyed and disappear in one crossing time, or in $\sim 1$ My. However, the coherent alignment of the clumps along the pillar indicates that they are long-lived structures, so that the velocity gradient is more naturally understood as due to rotation and spin. If the ETs are rotating at a velocity of $\sim 1-2$ \kms, the rotation period is $2 \pi r {\rm pc}/1.4 {\rm km\ s^{-1}} \sim 0.6$ My. 

Rotation directions of Pillars I and II are parallel at the same latitude. On the other hand, the spin direction is not systematic along each tail, as shown in the LV channel maps (figures \ref{lv_ch}), where the velocity gradient directions vary with latitude along the pillar. 
Moreover, the velocity field (moment 1) map in figure \ref{co_m12} shows varying gradient in the ETs.

These facts indicate that the rotation (spin) axes of the molecular clumps are random, suggesting that the clump rotation is originated in the clouds with turbulent motion before evolving into ETs. Such random rotation is in contrast to the regulated rotation and helical structure along the elephant trunk in the Rosette Nebula (Carlrqvist et al. 2002; Gahm et al. 2006).

\subsection{Magnetic confinement of gravitationally unbound clump}
\label{magconf}

In order for the DRI to dynamically compress the molecular clump, the momentum of the HII cap must be greater than, or at least comparable to that of the molecular clump. For this, the HII masses must be comparable because they are at rest, which is, however, not the case in such that the HII mass is two order of smaller than the molecular mass. Also, the external HII region of M16 surrounding the cap has no sufficient pressure. Hence, some other confinement mechanism is required.

We here propose that the magnetic field of the pillar can act as an effective container of the HII cap. As the conductivity of HII gas is sufficiently high, the gas is almost perfectly frozen into the magnetic field. This means that the HII cap can be confined near the molecular surface by the magnetic tension. In fact, the specific magnetic tension, which is on the order of $B^2/4 \pi \sim 3\times 10^{-9}$ erg cm$^{-3}$, is greater than the gaseous pressure in the cap of $\nel k \Te \sim 2\times 10^{-9}$ erg cm$^{-3}$. Note that this pressure is approximately equal to that of the dynamical pressure by the velocity dispersion in the molecular clumps.

The external compression is also necessary in order for the molecular clumps to be bound against the expansion due to the velocity dispersion and rotation. In fact, the virial mass is estimated to be $M_{\rm vir}\sim r \sigma_v^2/G \sim 10^2 \Msun$ for the observed radius $r\sim 0.14-0.18$ pc and velocity dispersion $\sigma_v\sim 1.7$ \kms in the tips of Pillars West I and East, which is much greater than the observed molecular masses from the CO line measurements (table \ref{tab_pillars}). Higher virial mass than the mass on the LTE assumption of CO lines has been also reported in another ET clumps in M16 (Andersen et al. 2004).  Similarly, comparable, or even greater mass is required to gravitationally bind the clumps against the centrifugal force by the rotation.

Thus, an external confinement such as due to the magnetic pressure and tension is necessary not only to maintain the high pressure of dense ionized gas on the clump's surface, assisting the RDI to work and making the pillar creative in star formation, but also to keep the molecular clump to be a bound system. However, there remains a question if such a non-virialized, hence gravitationally stable, clump can be a star forming site.

\section{Summary}

We investigated the molecular and radio continuum properties of the elephant trunks, or the Pillars of Creation, associated with the HII region M16 by analyzing the archival data of the CO line emissions from the FUGIN survey with the Nobeyama 45-m telescope and radio continuum survey of the Galactic plane surveys with the VLA. 

We showed that the head clump of Pillar  I composes the brightest radio source of thermal (free-free) emission in the entire M16 HII region. The radio source is categorized as a compact HII region of relatively small size and low density for its radio intensity and extent. The head clumps of the other ETs are also found to be strong radio sources. The radio morphology around the ETs is characterized by the cometary structure concave to the molecular trunk head. 

Although Pillar I's head is bright in radio emission, there exists no cataloged OB stars inside the pillar responsible for the heating of the HII gas. Instead, the radio luminosity of Pillar I head can be explained as due to the UV illumination by the nearest O5V star of NGC 6611. The radio intensity distribution around Pillar East head indicates a linear relation of the intensity against the injection-angle parameter, proving that the heating source is the same O stars in NGC 6611.

CO line kinematics revealed that the clumps in the pillars are rotating at $\sim 1-2$ \kms, comparable to the velocity dispersion and \Alf velocity. The spin directions are random, suggesting that the rotation is relic of turbulent cloud motion before the growth into ET. We proposed an interstellar TMO mechanism to explain the random rotational motion.

\vskip 2mm
{\bf Aknowledgements} The CO data were taken from the archives of the FUGIN Galactic plane survey using the Nobeyama 45-m telescope.
The 8 \mum, 20cm and 90 cm data were extracted from the archival data base MAGPIS (https://third.ucllnl.org/gps/index.html).
The data analysis was carried out at the Astronomy Data Center of the National Astronomical Observatory of Japan. 


\begin{thebibliography}{} 
 
 \bibitem[Andersen et al.(2004)]{2004A&A...414..969A} Andersen, M., Knude, J., Reipurth, B., et al.\ 2004, AA, 414, 969
 
\bibitem[Bertoldi(1989)]{1989ApJ...346..735B} Bertoldi, F.\ 1989, ApJ, 346, 735
 
\bibitem[\protect\citeauthoryear{Bisnovatyi-Kogan}{2007}]{2007MNRAS.376..457B} Bisnovatyi-Kogan G.~S., 2007, MNRAS, 376, 457 


\bibitem[\protect\citeauthoryear{Bolatto, Wolfire, \& Leroy}{2013}]{2013ARA&A..51..207B} Bolatto A.~D., Wolfire M., Leroy A.~K., 2013, ARA\&A, 51, 207 

\bibitem[\protect\citeauthoryear{Carlqvist, Gahm, \& Kristen}{2002}]{2002Ap&SS.280..405C} Carlqvist P., Gahm G.~F., Kristen H., 2002, Ap\&SS, 280, 405 
 


\bibitem[\protect\citeauthoryear{Chandrasekhar \& Fermi}{1953}]{1953ApJ...118..113C} Chandrasekhar S., Fermi E., 1953, ApJ, 118, 113 

\bibitem[\protect\citeauthoryear{Chauhan et al.}{2011}]
CChauhan N., Ogura K., Pandey A.~K., Samal M.~R., Bhatt B.~C., 2011, PASJ, 63, 795 

{\rev
\bibitem[Churchwell et al.(2009)]{2009PASP..121..213C} Churchwell, E., Babler, B.~L., Meade, M.~R., et al.\ 2009, PASP, 121, 213
}
 
\bibitem[\protect\citeauthoryear{Deharveng \& Maucherat}{1978}]{1978A&A....70...19D} Deharveng L., Maucherat M., 1978, A\&A, 70, 19 

\bibitem[Downes et al.(1980)]{1980A&AS...40..379D} Downes, D., Wilson, T.~L., Bieging, J., \& Wink, J.\ 1980, AA Suppl. 40, 379  
 

\bibitem[\protect\citeauthoryear{Ercolano et al.}{2012}]{2012MNRAS.420..141E} Ercolano B., Dale J.~E., Gritschneder M., Westmoquette M., 2012, MNRAS, 420, 141 


\bibitem[\protect\citeauthoryear{Evans et al.}{2005}]{2005A&A...437..467E} Evans C.~J., et al., 2005, A\&A, 437, 467 


\bibitem[\protect\citeauthoryear{Foster et al.}{2012}]{2012ApJ...756..128F} Foster A.~R., Ji L., Smith R.~K., Brickhouse N.~S., 2012, ApJ, 756, 128 
 

\bibitem[\protect\citeauthoryear{Frieman}{1954}]{1954ApJ...120...18F} Frieman E.~A., 1954, ApJ, 120, 18 
 

\bibitem[\protect\citeauthoryear{Gahm et al.}{2006}]{2006A&A...454..201G} Gahm G.~F., Carlqvist P., Johansson L.~E.~B., Nikoli{\'c} S., 2006, A\&A, 454, 201 
  

\bibitem[\protect\citeauthoryear{Gahm et al.}{2013}]{2013A&A...555A..57G} Gahm G.~F., Persson C.~M., M{\"a}kel{\"a} M.~M., Haikala L.~K., 2013, A\&A, 555, A57 

\bibitem[\protect\citeauthoryear{Getman et al.}{2012}]{2012MNRAS.426.2917G} Getman K.~V., Feigelson E.~D., Sicilia-Aguilar A., Broos P.~S., Kuhn M.~A., Garmire G.~P., 2012, MNRAS, 426, 2917 

\bibitem[\protect\citeauthoryear{Gonzalez-Alfonso \& Cernicharo}{1994}]{1994ApJ...430L.125G} Gonzalez-Alfonso E., Cernicharo J., 1994, ApJ, 430, L125 

\bibitem[\protect\citeauthoryear{Gritschneder et al.}{2010}]{2010ApJ...723..971G} Gritschneder M., Burkert A., Naab T., Walch S., 2010, ApJ, 723, 971 

 
\bibitem[Gritschneder et al.(2010)]{2010ApJ...723..971G} Gritschneder, M., Burkert, A., Naab, T., et al.\ 2010, ApJ, 723, 971

\bibitem[\protect\citeauthoryear{Guarcello et al.}{2007}]{2007A&A...462..245G} Guarcello M.~G., Prisinzano L., Micela G., Damiani F., Peres G., Sciortino S., 2007, A\&A, 462, 245 


\bibitem[\protect\citeauthoryear{Haikala et al.}{2017}]{2017A&A...602A..61H} Haikala L.~K., Gahm G.~F., Grenman T., M{\"a}kel{\"a} M.~M., Persson C.~M., 2017, A\&A, 602, A61 

\bibitem[\protect\citeauthoryear{Handa et al.}{1987}]{1987PASJ...39..709H} Handa T., Sofue Y., Nakai N., Hirabayashi H., Inoue M., 1987, PASJ, 39, 709 

\bibitem[\protect\citeauthoryear{Haworth \& Harries}{2012}]{2012MNRAS.420..562H} Haworth T.~J., Harries T.~J., 2012, MNRAS, 420, 562 

\bibitem[Helfand et al.(2006)]{2006AJ....131.2525H} Helfand, D.~J., Becker, R.~H., White, R.~L., et al.\ 2006, AJ, 131, 2525

\bibitem[\protect\citeauthoryear{Hester et al.}{1996}]{1996AJ....111.2349H} Hester J.~J., et al., 1996, AJ, 111, 2349 

\bibitem[Hill et al.(2012)]{2012A&A...542A.114H} Hill, T., Motte, F., Didelon, P., et al.\ 2012, AA, 542, A114

\bibitem[\protect\citeauthoryear{Hillenbrand et al.}{1993}]{1993AJ....106.1906H} Hillenbrand L.~A., Massey P., Strom S.~E., Merrill K.~M., 1993, AJ, 106, 1906 


\bibitem[\protect\citeauthoryear{Kohno et al. }{2019}]{2019PASJ...tmp}  Kohno, M., et al.
2019, PASJ, submitted.  

\bibitem[\protect\citeauthoryear{Mackey \& Lim}{2010}]{2010MNRAS.403..714M} Mackey J., Lim A.~J., 2010, MNRAS, 403, 714 
 
\bibitem[\protect\citeauthoryear{M{\"a}kel{\"a}, Haikala, \& Gahm}{2017}]{2017A&A...605A..82M} M{\"a}kel{\"a} M.~M., Haikala L.~K., Gahm G.~F., 2017, A\&A, 605, A82 

\bibitem[\protect\citeauthoryear{Markova et al.}{2004}]{2004A&A...413..693M} Markova N., Puls J., Repolust T., Markov H., 2004, A\&A, 413, 693 

\bibitem[\protect\citeauthoryear{Massi, Brand, \& Felli}{1997}]{1997A&A...320..972M} Massi F., Brand J., Felli M., 1997, A\&A, 320, 972 
 
\bibitem[\protect\citeauthoryear{McCaughrean \& Andersen}{2002}]{2002A&A...389..513M} McCaughrean M.~J., Andersen M., 2002, A\&A, 389, 513 

 \bibitem[McLeod et al.(2015)]{2015MNRAS.450.1057M} McLeod, A.~F., Dale, J.~E., Ginsburg, A., et al.\ 2015, MNRAS, 450, 1057 

 \bibitem[Oliveira(2008)]{2008hsf2.book..599O} Oliveira, J.~M.\ 2008, Handbook of Star Forming Regions, Volume II, 599

\bibitem[\protect\citeauthoryear{Osterbrock}{1957}]{1957ApJ...125..622O} Osterbrock D.~E., 1957, ApJ, 125, 622 

\bibitem[\protect\citeauthoryear{Panwar et al.}{2019}]{2019AJ....157..112P} Panwar N., Samal M.~R., Pandey A.~K., Singh H.~P., Sharma S., 2019, AJ, 157, 112 

\bibitem[\protect\citeauthoryear{Pattle et al.}{2018}]{2018ApJ...860L...6P} Pattle K., et al., 2018, ApJ, 860, L6 
 
\bibitem[\protect\citeauthoryear{Pineda, Caselli, \& Goodman}{2008}]{2008ApJ...679..481P} Pineda J.~E., Caselli P., Goodman A.~A., 2008, ApJ, 679, 481 

\bibitem[\protect\citeauthoryear{Pilbratt et al.}{1998}]{1998A&A...333L...9P} Pilbratt G.~L., Altieri B., Blommaert J.~A.~D.~L., Fridlund C.~V.~M., Tauber J.~A., Kessler M.~F., 1998, A\&A, 333, L9 


\bibitem[\protect\citeauthoryear{Plumpton}{1957}]{1957ApJ...125..494P} Plumpton C., 1957, ApJ, 125, 494 

\bibitem[\protect\citeauthoryear{Pottasch}{1956}]{1956BAN....13...77P} Pottasch S.~R., 1956, BAN, 13, 77 

\bibitem[\protect\citeauthoryear{Pound}{1998}]{1998ApJ...493L.113P} Pound M.~W., 1998, ApJ, 493, L113 

\bibitem[\protect\citeauthoryear{Reich et al.}{1990}]{1990A&AS...85..633R} Reich W., Fuerst E., Reich P., Reif K., 1990, A\&AS, 85, 633 

\bibitem[\protect\citeauthoryear{Reich et al.}{1984}]{1984A&AS...58..197R} Reich W., Fuerst E., Haslam C.~G.~T., Steffen P., Reif K., 1984, A\&AS, 58, 197 

\bibitem[\protect\citeauthoryear{Reid \& Ho}{1985}]{1985ApJ...288L..17R} Reid M.~J., Ho P.~T.~P., 1985, ApJ, 288, L17 
 

\bibitem[\protect\citeauthoryear{Schneps, Ho, \& Barrett}{1980}]{1980ApJ...240...84S} Schneps M.~H., Ho P.~T.~P., Barrett A.~H., 1980, ApJ, 240, 84 

\bibitem[Schuller et al.(2006)]{2006A&A...454L..87S} Schuller, F., Leurini, S., Hieret, C., et al.\ 2006, AA, 454, L87
 

\bibitem[\protect\citeauthoryear{Sherwood \& Dachs}{1976}]{1976A&A....48..187S} Sherwood W.~A., Dachs J., 1976, A\&A, 48, 187 
    

\bibitem[\protect\citeauthoryear{Sofue et al.}{2018}]{2018PASJ..tmp..102S} Sofue Y., et al., 2018, PASJ, 

\bibitem[\protect\citeauthoryear{Sofue}{2019}]{} Sofue Y. 2019, PASJ in press 
 @@
\bibitem[\protect\citeauthoryear{Sofue \& Reich}{1979}]{1979A&AS...38..251S} Sofue Y., Reich W., 1979, A\&AS, 38, 251 

\bibitem[\protect\citeauthoryear{Spitzer}{1954}]{1954ApJ...120....1S} Spitzer L., Jr., 1954, ApJ, 120, 1 
 

\bibitem[\protect\citeauthoryear{Sugitani et al.}{2007}]{2007PASJ...59..507S} Sugitani K., et al., 2007, PASJ, 59, 507 
 
\bibitem[\protect\citeauthoryear{Thompson, Smith, \& Hester} {2002}]{2002ApJ...570..749T} Thompson R.~I., Smith B.~A., Hester J.~J., 2002, ApJ, 570, 749 

\bibitem[Umemoto et al.(2017)]{2017PASJ...69...78U} Umemoto, T., Minamidani, T., Kuno, N., et al.\ 2017, PASJ, 69, 78 



\bibitem[\protect\citeauthoryear{Urquhart et al.}{2006}]{2006A&A...450..625U} Urquhart J.~S., Thompson M.~A., Morgan L.~K., White G.~J., 2006, A\&A, 450, 625 

\bibitem[\protect\citeauthoryear{Urquhart et al.}{2003}]{2003A&A...409..193U} Urquhart J.~S., White G.~J., Pilbratt G.~L., Fridlund C.~V.~M., 2003, A\&A, 409, 193 

\bibitem[\protect\citeauthoryear{Whalen \& Norman}{2008}]{2008ApJ...672..287W} Whalen D.~J., Norman M.~L., 2008, ApJ, 672, 287 
 
\bibitem[\protect\citeauthoryear{White et al.}{1999}]{1999A&A...342..233W} White G.~J., et al., 1999, A\&A, 342, 233  

\bibitem[\protect\citeauthoryear{Woodward, Helfer, \& Pipher}{1985}]{1985A&A...147...84W} Woodward C.~E., Helfer H.~L., Pipher J.~L., 1985, A\&A, 147, 84 

\bibitem[\protect\citeauthoryear{Xu et al.}{2019}]{2019A&A...627A..27X} Xu J.-L., et al., 2019, A\&A, 627, A27 


\bibitem[\protect\citeauthoryear{Zhu et al.}{2015}]{2015ApJ...812...87Z} Zhu F.-Y., Zhu Q.-F., Li J., Zhang J.-S., Wang J.-Z., 2015, ApJ, 812, 87 

\end{thebibliography}
\end{document}